# Upscaling Tomographic Volumetric 3D Printing via Virtual Stitching of Coordinated Projections

Hossein Safari[1,2], S. Kaveh Hedayati[3], Aminul Islam[3], Yi Yang[1,2,4*]

[1] Department of Chemistry, Technical University of Denmark; 2800 Kongens Lyngby, Denmark.

[2] Center for Energy Resources Engineering, Technical University of Denmark; 2800 Kongens Lyngby, Denmark.

[3] Department of Civil and Mechanical Engineering, Technical University of Denmark; 2800 Kongens Lyngby, Denmark.

[4] PERFI Technologies ApS; Kemitorvet 207, 2800 Kongens Lyngby, Denmark

* To whom correspondence should be addressed (email: yi@perfi.dk)

## Abstract

Tomographic volumetric 3D printing (TVP) offers layer-free, rapid fabrication of objects with high design freedom, but is limited to relatively small curing volumes because of the optical constraints imposed by an assumed need for telecentricity. We present a method to virtually stitch multiple projections from different light sources to build a single workpiece. To avoid the built-in requirement for telecentricity and thus the need for an index matching vat, projections are produced by a graphics processing unit (GPU)-accelerated raytracing solver. The method accounts for non-ideal light propagation, including non-telecentricity, reflection, refraction, attenuation, as well as any output power mismatch among projectors. It also iteratively optimizes projections to homogenize dose buildup in a curing volume up to 12.5 cm in diameter. Using two off-the-shelf DLP projectors, we show a printable area of 138 × 104 mm$^2$, with a native 48 μm pixel resolution. The versatility of the method is further demonstrated by rapidly fabricating objects up to 78.2 mm in height and 70.9 mm lateral dimension in less than 4 minutes, using both cubic and cylindrical containers. While our demonstration uses two light sources, the method can easily generalize to multi-projector systems, thus providing a cost-effective methodology to upscale high speed volumetric 3D printing.

## Introduction

Tomographic volumetric 3D printing (TVP) is an emerging vat photopolymerization method that mimics tomographic reconstruction to rapidly produce structures with high design freedom. In a typical TVP process, structured two-dimensional (2D) light patterns are projected into a rotating volume of photosensitive resin from multiple angles, creating a three-dimensional (3D) energy distribution that solidifies the entire 3D geometry within minutes. Previous research underscores TVP's advantages over conventional vat photopolymerization methods such as stereolithography (SLA) or digital light processing (DLP), emphasizing its rapid speed, smooth layer-free surfaces, and reduced reliance on support structures (*1–3*).

As a volume-at-once method (*3*, *4*), TVP has seen rapid progress, focusing on applicable materials, print resolution and reproducibility. Early implementations of TVP borrowed algorithms from medical computed tomography (CT), essentially using a filtered back-projection or Radon transform to derive projections. Many pioneering setups have thus employed telecentric optics paired with refractive index-matching liquids to control print fidelity, aligning with CT-derived principles (*1*, *5–10*). A core assumption of Radon transform-based approaches is the idealized condition that all incident rays are perfectly collimated and propagate in straight lines. While this assumption holds true for X-ray imaging, it becomes problematic for the range of wavelengths typically employed in TVP (UV and visible light), the propagation of which are susceptible to various optical phenomena such as refraction, reflection, attenuation, beam divergence, and scattering. These non-ideal optical effects undermine the parallel-beam assumption central to standard reconstruction algorithms, and often deteriorates print quality.

To address these limitations, computational methods have been developed to optimize the 2D light patterns and minimize geometric distortions. Noteworthy examples are gradient descent optimization (*1*, *8*)' object-space model optimization (OSMO) (*11*), maximum likelihood-expectation maximization (*12*), and band-constraint *Lp*-norm minimization (*13*). Others have employed ray-optics analysis to refine pixel intensities for further fidelity improvements. For example, Loterie et al. (*2*) developed a camera feedback system for offline optimization of pixel intensities in low etendue illumination, improving the print resolution down to 80 μm for positive features. Orth et al. proposed a post-calculation mathematical correction to the Radon transform that accounts for non-telecentric illumination and the container's lensing effects (*14*). Recent works have implemented more advanced light transport models to accurately compensate for non-ideal light propagation. For example, Webber et al. (*15*) introduced a three-dimensional raytracing (3DRT) algorithm that directly traces rays from the digital micromirror device (DMD), compensating for refraction and beam non-telecentricity. This strategy enabled a threefold increase in achievable build height; however, the computational complexity was up to two orders of magnitude higher than Radon-based methods. More recently, Nicolet et al. (*16*) developed an inverse-rendering framework based on the Mitsuba renderer (*17*), demonstrating successful prints in scattering resins as well as in square vials (edge length ≈ 12.4 mm). Nevertheless, the computational overhead associated with pattern generation remains substantial, requiring approximately 47 minutes, significantly longer than the actual printing process, thus partially eliminating TVP's speed advantage. From the current body of work, three critical barriers could be identified. First, traditional methods inherited from X-ray CT are computationally expensive while insufficiently modeling the complex interactions between light and resin. Second, while raytracing captures physical light-material interactions with high accuracy, it is even more computationally demanding. Third, many existing implementations rely on external libraries and rendering engines that can be difficult to adapt for specialized TVP settings. Therefore, there is a need for a computational framework tailored to the upscaling requirements of TVP, which can precisely capture light-material interactions, be computationally efficient, and be adaptive to various deployments of TVP apparatus.

An especially attractive aspect of TVP is that its build time remains largely independent of geometric complexity. Print speed is driven primarily by the resin's reaction kinetics and the projector's output power, rather than on the intricacy of workpiece's shape (*5*, *7*, *18*). Although numerous studies have explored computational optimization of exposure patterns (*1*, *11–13*, *15*), improvements in optics (*2*, *14*, *19*), the exploration of new materials (*5*, *9*, *20–22*), and limited multiphysics modeling (*23*, *24*), the scalability of TVP, particularly to larger part volumes, remains underexplored. TVP faces an inherent trade-off between build volume, printing speed, and achievable resolution. The resolution is fundamentally limited by the optical system (e.g., the projector's pixel size and point spread function of the lens system) and physical factors like light scattering and diffusion of chemical species (*18*, *25*). While high-resolution prints (tens of microns) have been demonstrated, they are typically confined to small build volumes (a few cubic centimeters) (*2–4*, *21*, *26*). Table 1 provides a comprehensive summary of key technological advancements and achieved print dimensions in TVP. Most systems remain limited to millimeters to a few centimeters in maximum dimension. For example, Kelly et al. achieved lateral dimensions of ~55 mm (*4*), and Chen et al. used a 40 mm-diameter vessel for work with high-attenuation resins (*27*). A few methods have aimed to address the scalability challenges. Roll-to-roll (R2R), for instance, unwraps the resin into a film, enabling the continuous production of extended, aperiodic structures with theoretically unlimited length (*9*). However, it currently relies on thermally gelling photoresists, and scaling up to larger footprints makes the unrolling of resin increasingly complex. Boniface et al. (*6*) introduced volumetric helical additive manufacturing (VHAM), which uses a helical motion to build taller parts. While this strategy does expand the build envelope vertically, total fabrication time scales directly with part height, reaching about 10 minutes for a 3 cm × 3 cm × 5 cm print.

Despite these efforts, scaling TVP to larger volumes remains constrained by fundamental light-transport and projection-optics limitations. As print volume increases, light must penetrate deeper into the resin, leading to additional attenuation from absorption and potentially scattering. In a single projector setup, this produces depth-dependent dose non-uniformity and under-exposure at voxels far from the source, effectively limiting the maximum printable area to only a few centimeters (see Table 1). Scaling also stresses the projection optics as the system must deliver sufficient irradiance while preserving spatial resolution over a larger field of projection (FOP). With fixed projector radiance and pixel pitch, enlarging the FOP spreads flux and relaxes the effective numerical aperture, degrading both dose rate and resolvable feature size. Although workarounds like helical movement expand the vertical FOP while preserving the optical resolution but it is at the direct expense of fabrication time for larger workpieces.

In this work, we address the fundamental challenges of upscaling TVP while preserving native optical resolution by combining multiple off-the-shelf projectors to effectively expand the FOP. Since the use of refractive index matching box becomes more cumbersome as the size of resin container increases or the shape of it changes, we developed original raytracing based projection generator that eliminates the need for index-matching vats and allows for flexible projector placement for large resin containers (up to Ø12.5 cm), compensating for volumetric attenuation as well as other non-ideal light propagation, and therefore expanding both vertical and lateral build dimensions. Accelerated by graphics processing units (GPU), our software is capable of rapidly calculating physically realistic light patterns under non-telecentric conditions and across diverse container geometries. By distributing the total dose from multiple overlapping projectors, we relax single-projector constraints on resolution and FOP, while providing a scalable platform for larger workpieces. We detail the theoretical foundation of this multi-projector system, evaluate its computational efficiency, and demonstrate its capability through fabricated examples. We conclude by discussing the influences of non-ideal optical effects, mechanical challenges and practical resolution limits.

**Table 1: Technological advancements and milestones in TVP**

| Study (Year) | Achievements | Emission wavelength (nm) | Computational methods | Materials used | Container shape and size | Reported max print dimension (mm) | Reported printing time (s) | Smallest feature size (µm) |
|---|---|---|---|---|---|---|---|---|
| Shusteff et al. (2017) (*28*) | Introduced Holographic 3D fabrication via multi-beam superposition | 532 | Phase-only hologram generation with Gerchberg-Saxton | Key precursor: poly(ethylene glycol) diacrylate Photoinitiator: Irgacure 784 | Cuvette (internal cross-section: 10 mm × 10 mm) | | ~1-10 | 300 - 400 |
| Kelly et al. (2019) (*1*) | Introduced Computed Axial Lithography (CAL) and demonstrated centimeter scale prints in acrylate polymer | 455 | Pattern generation: Radon Transform Optimization: Iterative least-squares | Key precursors: Bisphenol A glycerolate (1 glycerol/phenol) diacrylate Poly(ethylene glycol) diacrylate Photoinitiators: Camphorquinone (CQ) Ethyl 4-dimenthylamino benzoate (EDAB) | Cylindrical | Lateral: ~55 | ~30-300 | 300 |
| Bernal et al. (2019) (*20*) | Introduced Volumetric Bioprinting (VBP) generating cell-laden tissue constructs with high viability (>85%) from gelatin-based photoresponsive hydrogels | 405 | Pattern generation: Radon Transform | Key precursor: gelatin methacryloyl (gelMA) Photoinitiator: lithium phenyl(2,4,6-trimethylbenzoyl)phosphinate (LAP) | Cylindrical (diameter: 16.75mm) | ~ 5 | 22.7 | 145 |
| Cook et al. (2020) (*26*) | Used Volumetric Sdditive Manufacturing (VAM) to print in thiol-ene resins | 405 | Pattern generation: Radon Transform Optimization: Simultaneous algebraic reconstruction technique (SART) algorithm | Key precursors: Poly(ethylene glycol) diacrylate (Mn 250) (TEG-DA) Tris[2-(acryloyloxy)ethyl] isocyanurate (TAE-ICN) 1,3,5-triallyl-1,3,5-triazine-2,4,6(1H,3H,5H)-trione (TA-ICN) Photoinitiator: 2-methyl-4′-(methylthio)-2-morpholinopropiophenone or Irgacure 907 Additive: TEMPO | Cylindrical (diameter: 25 mm) | ~ 15 | | 850 |
| Loterie et al. (2020) (*2*) | Improved the feature resolution down to 80µm for positive features by introducing a feedback correction loop | 405 | Pattern generation: Radon Transform Optimization: Pattern correction with integrated feedback system | Key precursor: di-pentaerythritol pentaacrylate Photoinitiator: phenylbis(2,4,6-trimethylbenzoyl)phosphine oxide | Cylindrical (diameter: 17.5 mm) | ~ 16 | 20 | Positive: 80 Negative: 500 |
| Orth et al. (2021) (*14*) | Introduced a Radon-space resampling method to correct for non-telecentricity and vial refractions | 460 | Pattern generation: Filtered Radon Transform Optimization: Radon space resampling for ray distortion and nontelecentricity correction | Key precursors: bisphenol A glycerolate (1 glycerol/phenol) diacrylate poly(ethylene glycol) diacrylate Mn 250 g/mol Photoinitiators: camphorquinone ethyl 4-dimethylaminobenzoate | Cylindrical (nominal diameter :25.4 mm) | | 144 - 288 | |
| Bhattacharya et al. (2021) (*8*) | Introduced Penalty Minimization (PM) optimization method to improve the accuracy of light patterns | 405 | Pattern generation: Radon Transform Optimization: PM | Key precursor: urethane dimethacrylate IPDI Photoinitiator: 2-Methyl-4-(methylthio)-2-morpholinopropiophenone | Cylindrical (diameter :20 mm) | | | |
| Rizzo et al. (2021) (*29*) | Introduced an optimized photoresin for volumetric bioprinting based on fast thiol–ene step-growth photoclick crosslinking | 405 | | Key precursors: GLE-NB GEL-MA<br>Photoinitiator: lithium phenyl-2,4,6-trimethylbenzoylphosphinate (LAP) | Cylindrical | | ~11 - 38 | |
| Rackson et al. (2021) (*11*) | Introduced object-space model optimization (OSMO) algorithm to improve the accuracy of light pattern | 405 | Pattern generation: Radon Transform<br>Optimization: OSMO | | Cylindrical | | | |
| Wang et al. (2022) (*22*) | Introduced Dual Color Tomographic Volumetric 3D Printing (DC-TVP) for stiffness control | UV: 365<br>Visible: 455 | Pattern generation: Radon Transform<br>Optimization: Iterative least-squares | Key precursors: Bisphenol A glycerolate (1 glycerol/phenol) diacrylate<br>Poly(ethylene glycol) diacrylate<br>3,4-epoxycyclohexylmethyl 3,4-epoxycyclohexanecarboxylate<br>Photoinitiators: Camphorquinone<br>Ethyl 4-dimenthylamino benzoate<br>Triarylsulfonium hexafluoroantimonate salts | Cylindrical (diameter :15.5 mm) | | ~300-690 | |
| Bernal et al. (2022) (*5*) | Introduced the volumetric bioprinting of complex organoid-laden constructs, capturing key functions of the human liver | 405 | | Key precursor: gelMA<br>Photoinitiator: Lithium phenyl(2,4,6-trimethylbenzoyl)phosphinate (LAP) | Cylindrical (diameter: 10 mm) | ~8 | 19.5 | 42 |
| Toombs et al. (2022) (*21*) | Introduced microscale computed axial lithography (micro-CAL) of fused silica components | 442 | Pattern generation: Radon Transform Optimization: object-space model optimization | Key precursors: Trimethylolpropane ethoxylate triacrylate Hydroxyethylmethacrylate Silica glass nanocomposite Pentaerythritol tetraacrylate Photoinitiators: Camphorquinone Ethyl 4-dimenthylamino benzoate Irgacure 369 {2,2,6,6-tetramethyl-1-piperidinyloxy} | Cylindrical | 4 | | 20 |
| Kollep et al. (2022) (*30*) | Used tomographic volumetric printing to fabricate complex 3D centimeter-scale ceramic parts | 405 | Pattern generation: Radon Transform Optimization: Frequency filtering | Key precursors: Polysiloxane substituted precursor 1,4-butandiol diacrylate Photoinitiator: Diphenyl (2,4,6-trimethylbenzoyl) phosphin oxide | Cylindrical (diameter: 16.5 mm) | Lateral: ~16 | 60 | 200 |
| Madrid-Wolff et al. (2022) (*18*) | Introduced methods to control light scattering tomographic volumetric printing | 405 | Pattern generation: Radon Transform Optimization: Frequency filtering Gradient descent | Key precursors: Di-pentaerythritol pentaacrylate phenylbis (2,4,6-trimethylbenzoyl) phosphine oxide TiO2 nanoparticles Gelatin methacryloyl with cells Photoinitiator: Lithium phenyl-2,4,6-trimethylbenzoylphosphinate (LAP) | Cylindrical (diameter: 16 mm) | ~8 | 36 | |
| Rackson et al. (2022) (*31*) | Introduced a method for mitigating striations via a uniform optical exposure | 405<br>442 | | Key precursors: Bisphenol A glycerolate (1 glycerol/phenol) diacrylate<br>Poly(ethylene glycol) diacrylate<br>Pentaerythritol tetraacrylate<br>Photoinitiators: Irgacure 907<br>Camphorquinone<br>Ethyl 4-dimenthylamino benzoate | Cylindrical | | | |
| Orth et al. (2022) (*19*) | Introduced optical scattering tomography to monitor the real-time photopolymerization | Blue | Pattern generation: Filtered Radon Transform<br>Optimization: Radon space resampling for ray distortion and nontelecentricity correction | Key precursors: Diurethane dimethacrylate<br>Poly(ethylene glycol) diacrylate<br>Bisphenol A glycerolate (1 glycerol/phenol) diacrylate<br>Photoinitiators: Camphorquinone<br>Ethyl 4-dimenthylamino benzoate | Cylindrical (diameter :25 mm) | lateral: 14mm | | |
| Boniface et al. (2023) (*6*) | Introduced Volumetric Helical Additive Manufacturing (VHAM) to print taller structures | 405 | Pattern generation: Radon Transform<br>Optimization: Frequency filtering<br>Gradient descent | Key precursors: pentaacrylate commercial resin<br>Photoinitiator: phenylbis (2, 4, 6-trimethylbenzoyl) phosphine oxide | Cylindrical (diameter :32 mm) | Vertical: ~60<br>Lateral: ~30 | ~600 | 180 |
| Orth et al. (2023) (*25*) | Introduced deconvolution volumetric additive manufacturing correcting for the effect of diffusion and point-spread function (PSF) | 405 | Pattern generation: Radon Transform<br>Optimization: Frequency filtering<br>Deconvolution<br>Radon space resampling for ray distortion and nontelecentricity correction | Key precursor: Diurethane dimethacrylate<br>Poly(ethylene glycol) diacrylate<br>Photoinitiator: Ethyl (2,4,6-trimethylbenzoyl) phenylphosphinate | Cylindrical | lateral: 10mm | | 195 |
| Falandt et al. (2023) (*32*) | Introduced spatial-selective volumetric 4D printing and single-photon grafting of biomolecules within centimeter-scale hydrogels | 405 | | Key precursor: Commercial grade gelNOR (type B, bovine hide, DoF 60%) Photoinitiator: Lithium phenyl (2,4,6-trimethylbenzoyl) phosphinate | Cylindrical | 169.85 $mm^3$ | 20 | |
| Weisgraber et al. (2023) (*24*) | Introduced VirtualVAM: a simulation framework for modeling the tomographic VAM process | 405 | Pattern generation: Livermore Tomography Tools package (LTT) | Key precursors: Bisphenol A glycerolate (1 glycerol/phenol) diacrylate<br>Poly(ethylene glycol) diacrylate<br>Photoinitiator: Irgacure 907 | Cylindrical (diameter: 25.4 mm) | | 850 | |
| Thijssen et al. (2023) [NO_PRINTED_FORM] (*33*) | Introduced photo-cross-linkable poly(ε-caprolactone) networks through orthogonal thiol-ene chemistry | 442 | Pattern generation: Radon Transform<br>Optimization: object-space model optimization | Key precursor: Thiol-ene photocrosslinkable poly (ε-caprolactone)<br>Photoinitiator: Ethyl (2,4,6-trimethylbenzoyl) phenylphosphinate | Cylindrical | | | 100 |
| Chansoria et al. (2023) (*34*) | Introduced new algorithmic approaches for multi-material volumetric printing of complex shapes | 405 | Pattern generation: Radon Transform<br>Optimization: Frequency filtering<br>Gradient descent | Key precursor: Norbornene and thiol-modified gelatin<br>Photoinitiator: Lithium phenyl (2,4,6-trimethylbenzoyl) phosphinate | Cylindrical (diameter: 18 mm) | 10×10 $mm^2$ | | |
| Toombs et al. (2023) (*10*) | Introduced ethyl cellulose-based thermoreversible organogel photoresist for sedimentation-free volumetric additive manufacturing | 442 | Pattern generation: Radon Transform<br>Optimization: object-space model optimization | Key precursor: Trimethylolpropane triacrylate<br>Ethyl cellulose<br>Photoinitiators: Camphorquinone<br>Ethyl 4-dimenthylamino benzoate<br>{2,2,6,6,-tetramethyl-1-piperidinyloxy} | Cylindrical (diameter: 20 mm) | | | 283 |

| | | | | | | | |
|---|---|---|---|---|---|---|---|
| Webber et al. (2023) (*15*) | Introduced three-dimensional ray tracing (3DRT) correcting for nontelecentricity and vial refraction | 405 | Pattern generation: Ray tracing<br>Optimization: object-space model optimization | Key precursors: Diurethane dimenthacrylate<br>Poly(ethylene glycol) diacrylate<br>Photoinitiators: Camphorquinone<br>Ethyl 4-dimenthylamino benzoate<br>Ethyl (2,4,5-trimethylbenzoyl) phenylphosphinate | Cylindrical (diameter: 25 mm) | | |
| Xie et al. (2023) (*35*) | Used volumetric additive manufacturing in pristine silk-based (bio)inks | 525 | Pattern generation: Radon Transform<br>Optimization: Frequency filtering | Key precursor: Silk sericin and fibroin<br>Photoinitiator: Ruthenium/sodium persulfate | Cylindrical (diameter: 12 mm) | 57-228 | 57 |
| Barbera et al. (2024) (*36*) | Introduced volumetric printing of silica-based glasses with tunable multimaterial and microstructural control | 405 | Pattern generation: Radon Transform<br>Optimization: Frequency filtering<br>Gradient descent | Key precursors: Hydroxyethylmethacrylate<br>Trimethylolpropane ethoxylate triacrylate<br>Amorphous silica nanopowder<br>Diurethane dimethacrylate<br>Pentaerythritol tetraacrylate<br>Poly(diethoxysiloxane)<br>Triethyl borate<br>Photoinitiators: Diphenyl-(2,4,6-trimethylbenzoyl)-phosphinoxid<br>{2,2,6,6-tetramethyl-1-piperidinyloxy} | Cylindrical | | |
| Lian et al. (2024) (*37*) | Introduced rapid volumetric bioprinting of decellularized extracellular matrix bioinks | 525 | Pattern generation: Radon Transform<br>Optimization: Frequency filtering<br>Iterative optimization | Key precursors: Hydrogel with heart-derived decellularized extracellular matrix, and meniscus-derived decellularized extracellular matrix<br>Photoinitiators: Tris(2,2-bipyridyl) dichlororuthenium (II) hexahydrate<br>Sodium persulfate | Cylindrical (diameter: 6 mm) | 35 | 79 |
| Chen et al. (2024) (*27*) | Used expectation maximization algorithm to generate patterns for high-attenuation resins | 405 | Pattern generation: 3D exponential Radon Transform Optimization: Expectation maximization algorithm | Key precursors: Dipentaerythritol pentaacrylate Poly(ethylene glycol) diacrylate Dipentaerythritol pentaacrylate Tricyclo [5.2.1.02,6] decanedimethanol diacrylate Pentaerythritol triacrylate Photoinitiator: Irgacure 819 | Cylindrical (diameter: 40 mm) | | |
| Toombs et al (2024) (*9*) | Introduced Roll-to.roll tomographic additive manufacturing | 442 | Pattern generation: Ray tracing Optimization: Band-constraint Lp norm minimization | Key precursors: Trimethylolpropane triacrylate Ethyl cellulose Photoinitiators: Camphorquinone Ethyl 4-dimenthylamino benzoate Ethyl (2,4,5-trimethylbenzoyl) phenylphosphinate | | 46.5 $mm^2$ | |
| Nicolet et al. (2024) (*16*) | Introduced inverse rendering for pattern generation in scattering resins | 405 | Pattern generation: Inverse rendering Optimization: object-space model optimization | Key precusrsor: Di-pentaerythritol pentaacrylate (SR399) Photoinitiator: Photoinitiator: Phenylbis(2,4,6-trimethyl-benzoyl)-phosphine oxide (TPO) | Cuvette (diameter : 12.43mm) Cylindrical (diameter : 16.77mm) | | 27-162 |

## Virtual stitching of projections

Building large workpieces using TVP while maintaining the projector's native pixel pitch requires in each projection the use of a very large number of pixels, often exceeding the total number of mirror units on a single digital micro-mirror device (DMD). It is intuitive to pool the pixels from multiple DMDs to deliver a single projection containing many pixels. For TVP, the DMDs do not have to be paneled side-by-side as is the case in the upscaling of DLP-based vat photopolymerization. By arranging multiple projectors in lateral and/or vertical configurations and virtually stitching their overlapping fields of view at the focal plane, one can form the high-resolution sinogram needed for printing large, complex designs. Fig. 1a highlights this challenge by presenting a 3D design whose sinogram for a single lateral slice contains 11,619 pixels, well beyond the capacity of a single projector. Mathematically, the sinogram $S(x, y, \varphi)$ encodes the source (projected) intensity for each pixel *(x,y)* throughout a full resin rotation $\varphi \in [0, 2\pi)$. Under idealized (no-absorption) condition, analogous to the Radon transform, integrating $S(x, y, \varphi)$ across $\varphi$ yields the cumulative source intensity that would reach a voxel at *(X,Y)*. However, in a physical system, the actual dose *D(X,Y)* received by a voxel is further modulated by attenuation: $D(X, Y) = \int_{\varphi=0}^{2\pi} S(x, y, \varphi) e^{-(\sigma \ell(X,Y,\varphi))} d\varphi$, where $\sigma$ is the resin's attenuation coefficient, and $\ell(X, Y, \varphi)$ is the path length from the projector's source plane to the voxel at *(X,Y)*. We first consider a hypothetical scenario in which the design is only extensive laterally, necessitating only lateral alignment and stitching of the projectors. Suppose we have *N* projectors (with *N* = 4 in Fig. 1b), each covering a finite two-dimensional region of the image plane. In an idealized, telecentric setting with parallel-beam projection geometry (assuming no refraction at the air-container interface), one can treat each projector $p \in \{1, \dots, N\}$ as projecting onto a planar region $\Omega_p \subset \mathbb{R}^2$ as the subset of the total image domain $\Omega$ at the focal plane. The stitched intensity at angle $\varphi$ then becomes:

$I_{tot}(x,y,\varphi) = \sum_{p=1}^{N} I_p(x,y,\varphi)\,\chi\Omega_p(x,y)$, where $\chi\Omega_p(x,y)$ is 1 if $(x,y) \in \Omega_p$ and 0 otherwise, and $I_p(x,y,\varphi)$ is the portion of the sinogram that belongs to projector $p$ (Fig. 1c). To enhance lateral coverage beyond the cumulative native resolution of individual projectors, projectors could be oriented diagonally or positioned off-center, effectively increasing the total pixel count available for fabricating large-scale structures. Fig. 2a illustrates a more realistic scenario, where multiple projectors are arranged both laterally and vertically to collectively cover the entire image domain. In this configuration, projectors 1 and 3 primarily illuminate the central region, while projectors 4 and 2 cover the upper and lower regions, respectively. Fig. 2b compares example sinograms projected by each projector (S1, S2, and S3) contrasting scenarios with and without angular compensation. The uncompensated sinograms assume an identical viewing angle for all projectors, which is physically impractical. Normally, each projector has a unique angular offset $\theta_p$, resulting in distinct effective viewing angles. If the resin container rotates at a nominally constant angular velocity, defining the rotation angle at time $t$ as $\varphi(t)$, the effective viewing angle for projector $p$ at time t becomes $\alpha_p(t) = \varphi(t) - \theta_p$. Therefore, projector $p$ displays its sub-sinogram based on its effective viewing angle $\alpha_p(t)$ . An important practical consideration, not inherently addressed during sinogram generation, is the variation in output power among projectors, which must be managed to achieve simultaneous and uniform polymerization of the workpiece. Because the correlation between a projector's output power and the gray values of its pixels can differ, calibration is required when computing projections. Fig. 2c illustrates example output-power correlations for the projectors depicted in Fig. 1b. To calibrate pixel intensities across projectors, one can designate the projector with the lowest output power as the reference and then scale all other projectors to match its power correlation. For instance, if projector 2 serves as the reference, the desired gray value $I_2'$ for a given intensity $I_2^0$ can be calculated using $I_2' = F_2^{-1}(G(I_2^0))$. This ensures that the light patterns from multiple projectors produce the same output power, thereby facilitating uniform dose delivery and consistent polymerization across the entire workpiece.

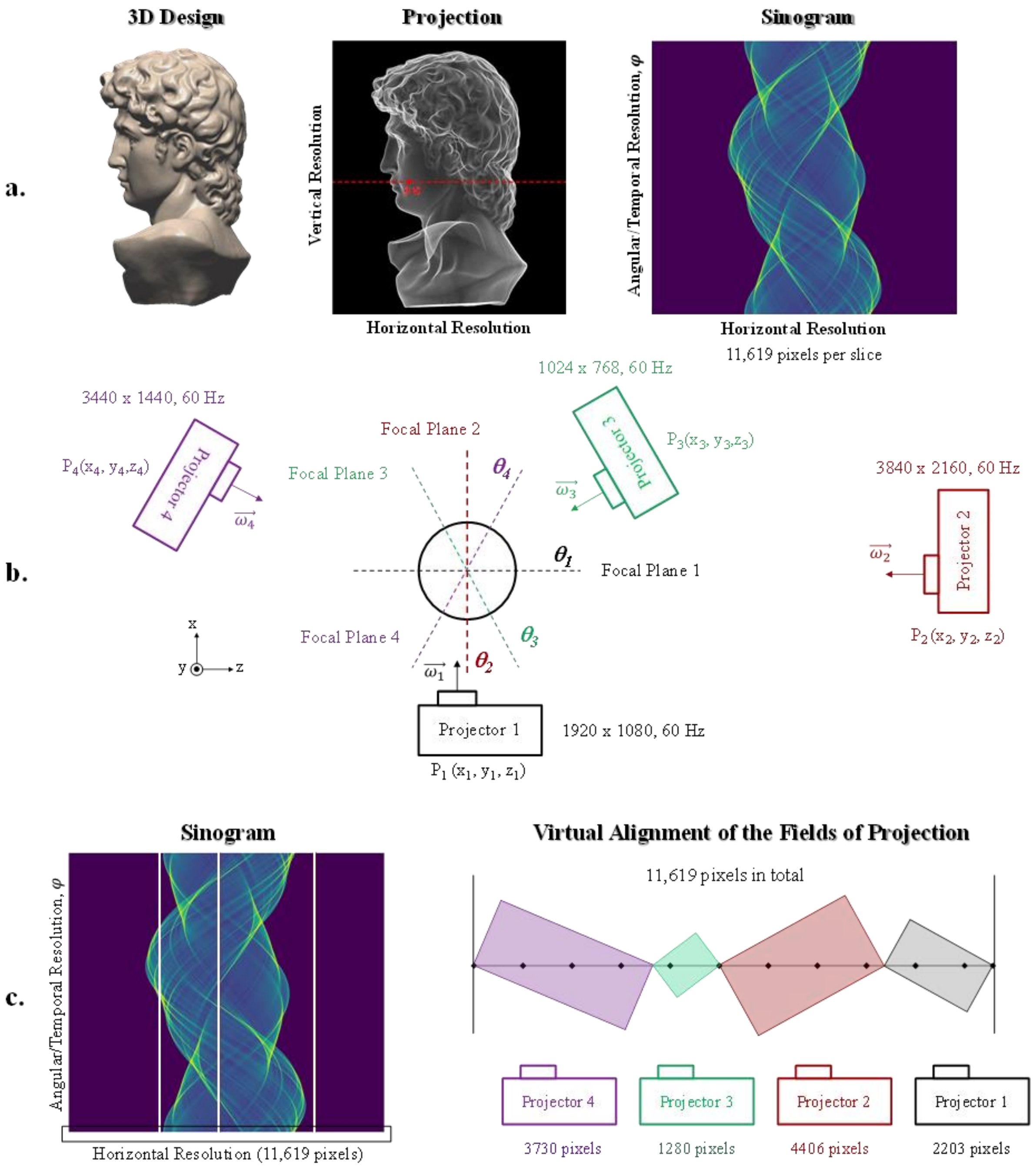

**Fig. 1: Virtual alignment and stitching of multiple projectors can deliver a high-resolution sinogram. (a)** A 3D design (left) and its 2D projection (middle), with a red dashed line marking a single slice spanning 11691 pixels. The sinogram (right) shows how projected intensities in this slice evolve over a full rotation period. (**b)** Schematic of a multi-projector arrangement in which four projectors, placed at arbitrary positions $P_i$ around the curing volume, each point in a direction $\omega_i$ and view the resin at an angle $\theta_i$ ($i\in\{1,2,3,4\}$). Focal planes are co-located with the rotation center to maintain proper focus, and each projector can operate at a single or multiple wavelengths depending on practical requirements. Notably, each projector has its own native resolution and working distance, with projectors oriented diagonally to maximize the horizon of projection. **(c)** Virtual stitching of the projectors' fields of view collectively covers the full sinogram. Diagonal orientation and off-centering of the projectors enable an effective resolution that exceeds any single projector's native capability.

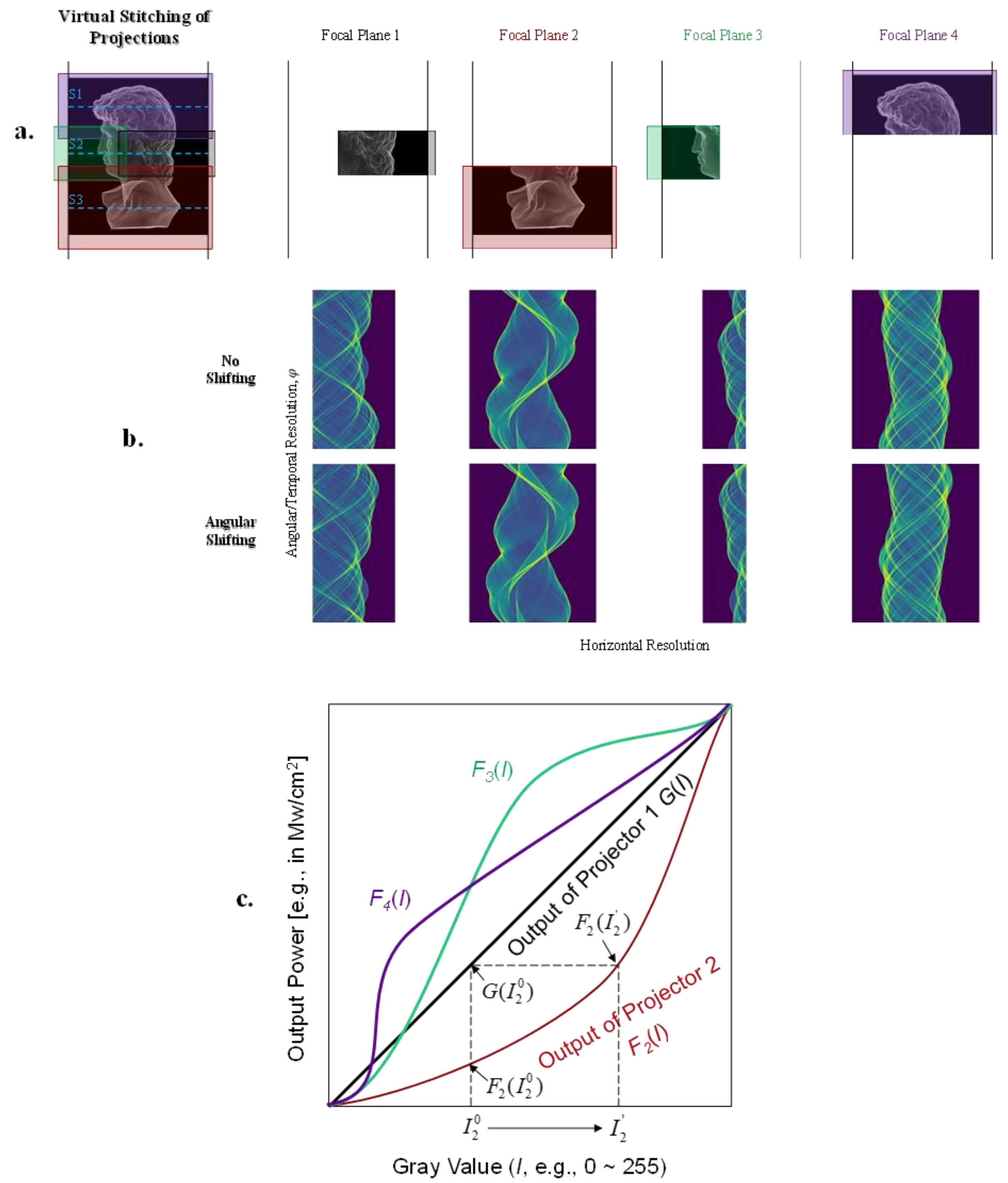

**Fig. 2: Physical alignment and intensity scaling of multiple projectors are essential for achieving a uniform dose distribution. (a)** Schematic illustration of the lateral and vertical projection alignments, showing how individual projector fields are stitched together to cover the entire imaging domain at the focal plane. **(b)** Shifting of projected sinograms (S1, S2, and S3) due to the spatial positions of individual projectors relative to the curing volume. Note that projector 2's sinogram remains unchanged since this projector is located at the reference angle (0°). **(c)** Calibration curves depicting exemplary variations in optical output power among different projectors. By choosing the projector with the lowest

power output (e.g., Projector 2) as the baseline reference, pixel intensities of the remaining projectors can be scaled accordingly.

## Raytracing-based projection generation

Accommodating large, arbitrarily shaped containers is key to upscaling TVP. To achieve this, the index-matching vat was avoided, and realistic physics models were adopted to describe light propagation. Traditional methods borrowed from X-ray CT are both computationally demanding and limited in capturing the complex interactions between the resin materials and the printing beam(s). By contrast, raytracing effectively handles these interactions. Although it can be computationally intensive, modern GPU architectures provide significant acceleration through parallel processing. Raytracing also allows the use of standard off-the-shelf light projectors, avoiding costly customary bi-telecentric set of optical lenses, making upscaling TVP more economical. Here, we present a framework designed to generate projections for multiple projectors surrounding a photoresin container of arbitrary shape. The framework is based on a rendering approach that uses a radiative transfer model to simulate the light propagation through participating 3D volumes.

**Light propagation** The key components involved in light propagation in a typical TVP setup include the light source, the DMD chip, optical lenses, a container, and the photopolymer precursors. A full optical simulation of every lens surface requires detailed manufacturing data, which is often unavailable when using off-the-shelf projectors. Therefore, we treated each projector as an assembly (a single effective optical module), capturing its cumulative impact on ray generation rather than modeling each internal lens surface. In practice, the positions and orientations of these generated rays are determined by several factors, such as the spatial arrangement of projectors around the curing volume, the configuration of the DMD, and the lens parameters (e.g., focal length, numerical aperture). Once generated, these rays are traversed until they hit the container's wall where refraction and reflection are modeled.

Cylindrical containers have been popular choices in TVP due to their simple, symmetric geometry, which facilitates more uniform illumination of the curing volume and simplifies ray-surface intersection calculations. By contrast, the use of arbitrarily shaped, non-cylindrical containers tailored to geometric characteristics of specific workpieces remains relatively unexplored, with only one study examining square cuvettes (*16*). Our ray tracer uses a polygon mesh representation of the container, integrated with a Bounding Volume Hierarchy (BVH) acceleration structure that enables fast and efficient ray-triangle intersections for contains of arbitrary size and shape (Fig. S4).

The reflection and refraction at the container's interface significantly influence the light dose distribution within the resin. To accurately model these interactions, we employ the Bidirectional Scattering Distribution Function (BSDF) (*38*), denoted $f_s(x, \vec{\omega}_i, \vec{\omega}_o)$. The BSDF describes the ratio of outgoing radiance $dL_o(x, \vec{\omega}_o)$ in a direction $\vec{\omega}_o$ to the incoming radiance $dE_i(x, \vec{\omega}_i)$ from direction $\vec{\omega}_i$ at a point $x$ on the surface:

$$f_s(x, \vec{\omega}_i, \vec{\omega}_o) = \frac{dL_o(x, \vec{\omega}_o)}{L_i(x, \vec{\omega}_i) cos\theta_i d\vec{\omega}_i} \quad (1)$$

where $L_i$ is the incoming radiance, $\theta_i$ is the angle of incidence with respect to the surface normal $\boldsymbol{n}$, and $d\vec{\omega}_i$ is the incident solid angle. The BSDF is composed of two main components: The Bidirectional Reflectance Distribution Function (BRDF), $f_r$, which describes light reflected from the surface, and the Bidirectional Transmittance Distribution Function (BTDF), $f_t$, which describes light transmitted through the surface.

For smooth interfaces, the reflections and refractions become purely specular, meaning light is scattered into discrete directions. For these surfaces, the partitioning of incident light energy into reflected and transmitted components at the container's surface is governed by the Fresnel equations. For unpolarized light, the Fresnel reflectance, $F_r$, which dictates the fraction of light reflected, is calculated as the average of its s-polarized, $R_s$, and p-polarized, $R_p$, components:

$$R_s = \left|\frac{n_1 \cos\theta_i - n_2 \cos\theta_t}{n_1 \cos\theta_i + n_2 \cos\theta_t}\right|^2 \tag{2}$$

$$R_p = \left|\frac{n_2 \cos\theta_i - n_1 \cos\theta_t}{n_2 \cos\theta_i + n_1 \cos\theta_t}\right|^2 \tag{3}$$

$$F_r = \frac{R_s + R_p}{2} \tag{4}$$

Here, $n_1$ and $n_2$ are the refractive indices of the two adjacent media (e.g., air and the container material); $\theta_i$ is the angle of incidence relative to the surface normal ***n***; and $\theta_t$ is the angle of transmission (refraction) given by Snell's law, $n_1 \sin(\theta_i) = n_2 \sin(\theta_t)$. Based on this, the specular BRDF and BTDF for a smooth surface are formulated as:

$$f_r(\vec{\omega}_i, \vec{\omega}_o) = F_r(\theta_i, n_1, n_2)\frac{\delta(\vec{\omega}_o - \vec{\omega}_r)}{cos\theta_r} \tag{5}$$

$$f_t(\vec{\omega}_i, \vec{\omega}_o) = T_p(\theta_i, n_1, n_2)\delta(\vec{\omega}_o - \vec{\omega}_t) \tag{6}$$

where $\delta(.)$ is the Dirac delta function, $\vec{\omega}_r$ is the mirror reflection of $\vec{\omega}_i$ about *n*, and $\vec{\omega}_t$ is the refracted (transmitted) direction determined by Snell's law. The transmitted lobe is weighted by the power transmittance coefficient:

$$T_p = (1 - F_r)\frac{n_2 cos\theta_t}{n_1 cos\theta_i} \tag{7}$$

which accounts for the change in projected solid angle and refractive index across the interface, ensuring energy conservation.

Total Internal Reflection (TIR) occurs if the incidence angle $\theta_i$ exceeds the critical angle $\theta_c$, where $\theta_c = sin^{-1}\left(\frac{n_2}{n_1}\right)$, when light attempts to pass from a denser medium to a less dense medium ($n_1 > n_2$). If TIR occurs, $F_r = 1$, all incident light is reflected according to the BRDF, and the BTDF term becomes zero, meaning no refraction takes place. Otherwise (if $\theta_i \leq \theta_c$ or if $n_1 \leq n_2$), the portion of light that enters the container continues as a transmitted (refracted) ray, following the direction dictated by the BTDF. This transmitted ray can then intersect additional internal surfaces, such as those of the photoresin, where reflection and refraction are again computed using the BSDF framework with appropriate Fresnel terms based on the refractive index of the material inside the container.

**Volume rendering** The photoresins used in TVP can be modelled as homogenous participating media consisting of particles that absorb or scatter incident illumination. The absorption and out-scattering attenuate the energy of the light beam as it passes through the resin medium, while in-scattering, and any volumetric emission if present, would contribute to the radiance along the light's path. The fundamental equation governing the change in radiance, $L(x,\vec{\omega})$, at position $x$, in direction $\vec{\omega}$, as light propagates through a participating medium is the Radiative Transfer Equation (RTE) (*39*):

$$\begin{aligned}(\vec{\omega}.\nabla)L(x,\vec{\omega}) = &-\sigma_t(x)L(x,\vec{\omega}) \\ &+ \sigma_s(x)\int_{S^2} p(x,\vec{\omega},\vec{\omega}')L(x,\vec{\omega}')d\omega' \\ &+ \epsilon(x,\vec{\omega})\end{aligned} \tag{8}$$

Here, $(\vec{\omega}.\nabla)L(x,\vec{\omega})$ denotes the directional derivative of radiance. The extinction coefficient $\sigma_t = \sigma_a + \sigma_s$ is the sum of absorption coefficient $\sigma_a$ and scattering coefficient $\sigma_s$. The phase function $p(x,\vec{\omega},\vec{\omega}')$ describes the angular distribution of scattered light, and $\epsilon(x,\vec{\omega})$ represents the volumetric emission term.

For applications involving large print volumes, ensuring that the writing light can propagate at least halfway through the material is crucial. Therefore, highly transparent (optically clear) resins are preferred to minimize attenuation due to absorption and scattering. In such resins, if volumetric scattering and emission effects are negligible (i.e. $\sigma_s \approx 0, \epsilon \approx 0$), the RTE equation can be approximated by:

$$(\vec{\omega}.\nabla)L(x,\vec{\omega}) = -\sigma_a(x)L(x,\vec{\omega}) \tag{9}$$

Integrating this simplified RTE for a light beam traveling from an initial position $x = 0$ to a position $x=s$ along the direction $\vec{\omega}$ yields:

$$L(s) = L(0)\exp\left(-\int_0^s \sigma_a(x)\, dx\right) \quad (10)$$

For example, if the absorbing medium is homogeneous and $\sigma_a$ is a constant, this expression reduces to the Beer-Lambert law:

$$L(s) = L(0)e^{-\sigma_a s} \quad (11)$$

Within our computational ray tracing framework, this exponential attenuation is quantified as a local light transmittance $T_j$ for each voxel j intersected by a ray:

$$T_j = e^{-\sigma_a \xi_j l_{ij}} \quad (12)$$

Where $\sigma_a$ is the absorption coefficient of the resin, $\xi_j$ is the modulating weight sampled from voxel $j$, and $l_{ij}$ is the path length of the ray $i$ within voxel $j$.

To efficiently simulate ray propagation through the discretized 3D target geometry (volume grid), the Digital Differential Analyzer (DDA) algorithm is employed (*40*). This method systematically identifies each voxel that a ray intersects, enabling accurate sampling of the volumetric properties along the ray's path. As each ray traverses the volume under DDA, the individual transmittances from all intersected voxels are compounded to determine the total attenuation experienced by the ray along its path. This cumulative transmittance $T_i$ for ray $i$ is calculated by multiplying the transmittances of all intersected voxels:

$$T_i = \prod_{j=1}^{N} T_j = \prod_{j=1}^{N} e^{-\sigma_a \xi_j l_{ij}} = e^{-\sigma_a \sum_{j=1}^{n} \xi_j l_{ij}} \quad (13)$$

where $N$ is the total number of voxels intersected by ray $i$.

At the exit point of the bounding box, which defines the spatial limits of the curing volume, the final transmittance $T_i$ represents the fraction of the initial light intensity that has not been attenuated after traversing the resin. The pixel intensity $I_i$ is then computed by subtracting this transmitted fraction from the initial (normalized) intensity:

$$I_i = 1 - T_i = 1 - e^{-\sigma \sum_{j=1}^{n} \xi_j l_{ij}} \quad (14)$$

## Results and discussion

Printing larger structures in TVP introduces an array of new obstacles. Rescaling conventional setups, which typically rely on inverted (top-down) rotation stages and index-matching baths, quickly becomes impractical once containers grow in size and weight. Upscaling requires new apparatus design along with a resin whose absorption length matches the container's radius to ensure adequate light penetration. We developed a minimalistic large-volume printing setup (Fig. 3) that addresses these requirements. It integrates two off-the-shelf laser projectors equipped with non-telecentric optical lenses to achieve focused light delivery into the resin.

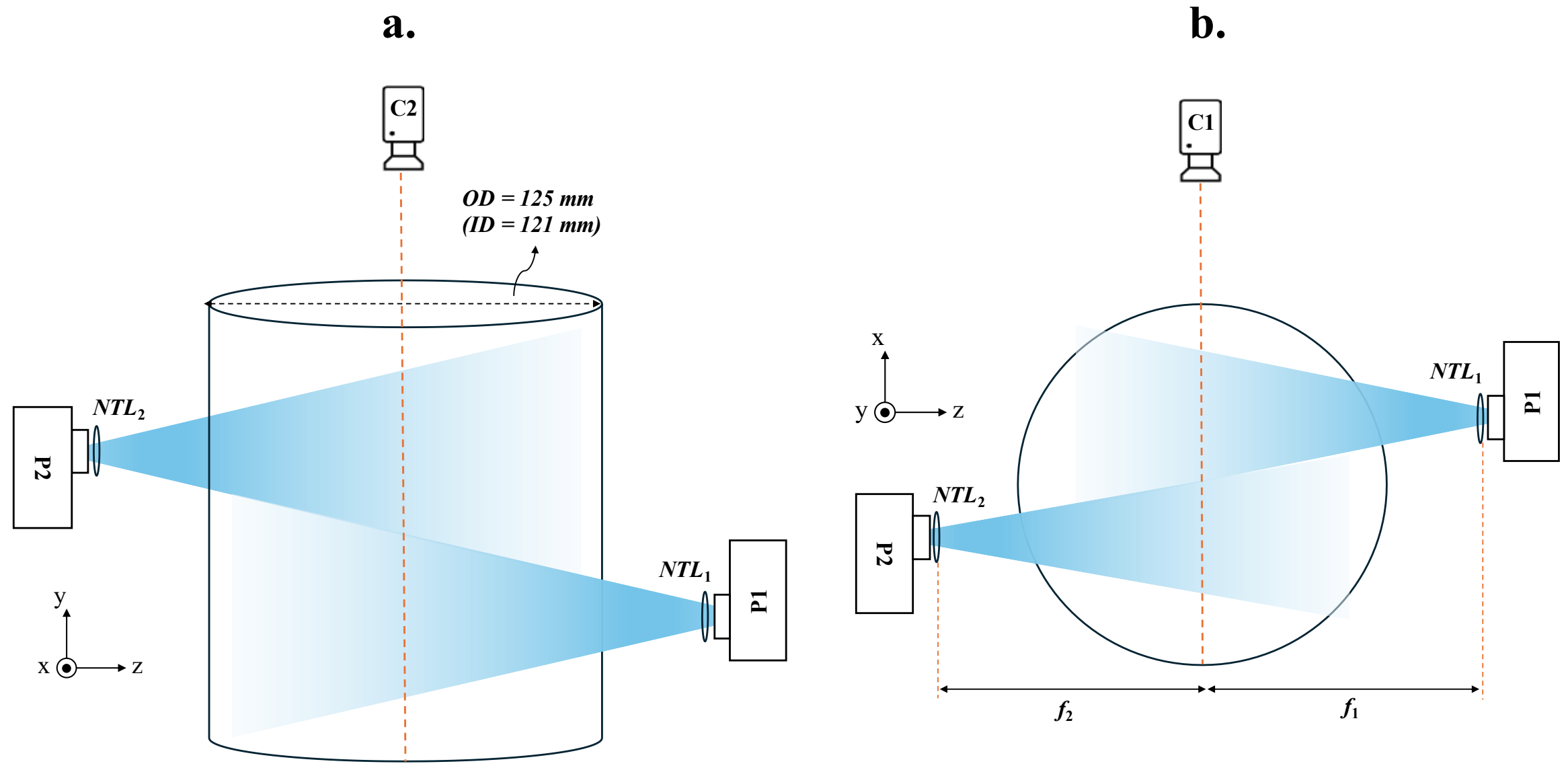

**Fig. 3: Schematic design of the dual-projector setup used in this study, (a)** side view and **(b)** top view. Each projector uses a non-telecentric lens ($NTL_1$ and $NTL_2$), characterized by focal lengths $f_1$ and $f_2$, respectively. Major system components are outlined in Table S1.

The system is designed to accommodate large photoresin containers while ensuring critical rotational stability. We demonstrate that by relying on readily available optical components and our in-house raytracing software. To prevent convection or sedimentation during printing, we blended Bisphenol A glycerolate (1 glycerol/phenol) diacrylate (BPAGDA) and poly(ethylene glycol) diacrylate (PEGDA) at a 2:1 ratio so that the resin is viscous enough to hold features steady, but not too viscous that it becomes impractical to handle during post-processing. UV-Vis spectroscopic analysis of the resin (Table S2) at various initiator concentrations indicated a primary absorbance peak around 470 nm (Fig. S5a), aligning well with the projectors' emission peak near 455 nm (Fig. S6a). These measurements helped tailor different resin formulations to specific container sizes. For example, when printing in a 125 mm-diameter cylindrical vessel, a combination of 0.025 wt.% CQ and 0.05 wt.% EDAB ensured sufficient volumetric coverage by the curing light. Although both projectors were from the same manufacturer and had similar emission profiles, their measured output powers differed slightly (Fig. S6b). Therefore, when both projectors were involved, we used the lower-powered projector as a reference and rescaled the second projector's intensity so that the entire volume would illuminate uniformly.

All ray-tracing computations were performed on a desktop workstation featuring an NVIDIA RTX 4090 GPU (24 GB VRAM) and 96 GB of system RAM. Although, our raytracing software is capable of handling both regular grid and octree representation of the 3D models, ample GPU memory allowed us to opt for the regular grid approach. While octree grids significantly reduced memory usage, we observed no speed advantage in our case. Using regular-grid voxelizations as GPU textures yielded a performance gain of two-to-three times over octree-based implementations for most of the tested geometries. Table 1 provides an overview of the container geometries, resolution parameters, and the ray tracer's computational performance for each 3D model used in this study.

**Table 1: Overview of container geometries, resolution settings, and computational performance for the 3D models used in this study**

| Model | Container shape | Container dimension ($mm^3$) | Pixel size (mm) | Voxel size (mm) | Number of pixels ×Angles | Number of voxels | Rays per pixel | Optimization time (s) |
|---|---|---|---|---|---|---|---|---|
| David | Cylindrical | 125×150×125 | 0.048 | 0.144 | 2880×2135 ×360 | 296×543×348 | 3 | 1125 |
| Francis of Assisi | Cylindrical | 50×150×50 | 0.048 | 0.096 | 2880×2060 ×360 | 261×752×206 | 3 | 780 |
| Ear | Cylindrical | 125×150×125 | 0.048 | 0.096 | 2880×1080 ×360 | 588×246×738 | 3 | 915 |
| David | Cubic | 74×74×74 | 0.048 | 0.096 | 1920×1080 ×360 | 334×547×332 | 3 | 375 |

**Dimensional accuracy**

Given the off-center placement and non-telecentric geometry of the projectors in our setup, it was important to confirm that our raytracing software can effectively compensate for various optical effects such as refraction and ray divergence (Fig. 3). To assess the accuracy of printing, we designed a custom pyramid-shaped calibration part composed of four distinct circular cross-sectional tiers (Fig. 4a, b). The largest cross-section measures 40 mm in diameter, and the smallest, located at the apex, measures 10 mm in diameter. A central negative feature with a 5 mm diameter was included to evaluate the system's dimensional fidelity in capturing internal geometries.

For validation, we used one of the projectors positioned off-axis to provide partial illumination, effectively projecting only half of the designed projection pattern. Despite this optical asymmetry and the intrinsic complexity of refractive effects (see Materials and Methods, Experimental Setup), our system accurately produced the calibration structure, capturing all designed positive and negative features (Fig. 4c). These results show the effectiveness of our raytracing approach in accounting for non-telecentric, high-etendue beam and for refractive distortions. Throughout the printing, real-time monitoring was conducted using a dark-field imaging system, which captured scattered illumination from the projector. The intensity of this scattered light correlated with the degree of resin polymerization, increasing notably toward the end of the curing process. The entire calibration part was fabricated in less than one minute, a noteworthy observation given the low photoinitiator concentration utilized in our resin formulation (0.025 wt.% CQ and 0.05 wt.% EDAB). During the printing, the central cross-sectional features appeared initially, followed shortly by near-simultaneous polymerization of the remaining sections. Minor deformation was caused during post-processing.

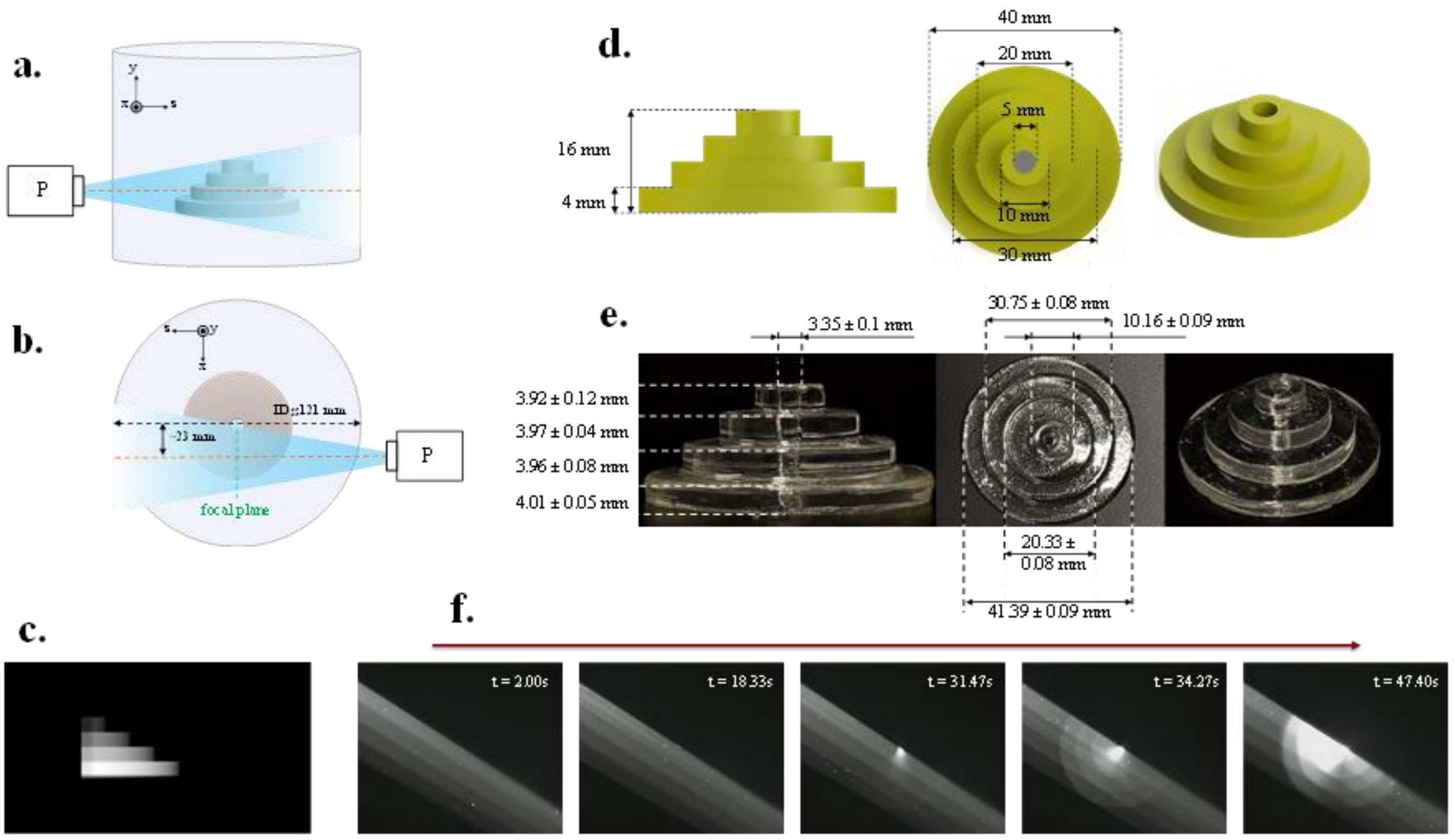

**Fig. 4: (a)** Side and **(b)** top view illustration of the custom part centrally positioned within a large-diameter (ID ≈ 121 mm) cylindrical container. The lateral offset is approximately 23 mm. **(c)** Calculated projection incorporating non-telecentricity, refraction, and projector positioning. **(d)** Multi-angle illustration of the part along with its design parameters. **(e)** Printed workpieces shown from various angles. **(f)** Top-view time-lapses of the printing process captured using a dark-field imaging system.

## Printing in non-cylindrical containers

To validate raytracing against surface irregularities, we extended our investigation to printing in non-cylindrical geometries. Printing objects in arbitrarily shaped containers requires an accurate 3D representation of the container's surfaces. For demonstration purpose, we selected a cubic vat (edge length 74 mm), and idealized the cube as having perfectly planar, mutually orthogonal faces and illuminated it with a single projector.

We printed a replica of the cast of Michelangelo's David (head, National Gallery of Denmark) measuring approximately $32 \times 52.5 \times 31.8$ mm$^3$. Fig. 5a illustrates a time-lapse of the printing process within the cubic container (see Mov. S2). Successful fabrication required generating light patterns that compensated for multiple optical effects, including beam non-telecentricity (visible in Fig. 5a), as well as surface reflections, and refraction. A special case of refraction happens when the container's edge is illuminated (see Fig. 5b,c), causing two adjacent faces to refract the beam at different angles. As a result, unlike cylindrical containers, the projection angle of the light patterns for the cubic container must align precisely with its orientation.

Fig. 5e shows the completed print, indicating successful formation of most features. However, the nose consistently failed to print completely despite multiple trials. We attribute this to discrepancies between our idealized cubic model and the actual container geometry, especially at the edges. Efforts to mitigate these artifacts by excluding projections near the edges resulted in undesirable rectilinear distortions without resolving the primary issue (Fig. 5f,g). Therefore, while ray tracing compensates effectively for optical path errors and helps dimensional accuracy, the intrinsic design of cubic containers combined with orthogonal projection of the light patterns restrict precise printing of fine surface details. A potentially more effective route would be to fix the vat and move the projector to multiple, non-orthogonal viewpoints, sacrificing strict tomographic reconstruction in favor of uniform, artefact-free exposure.

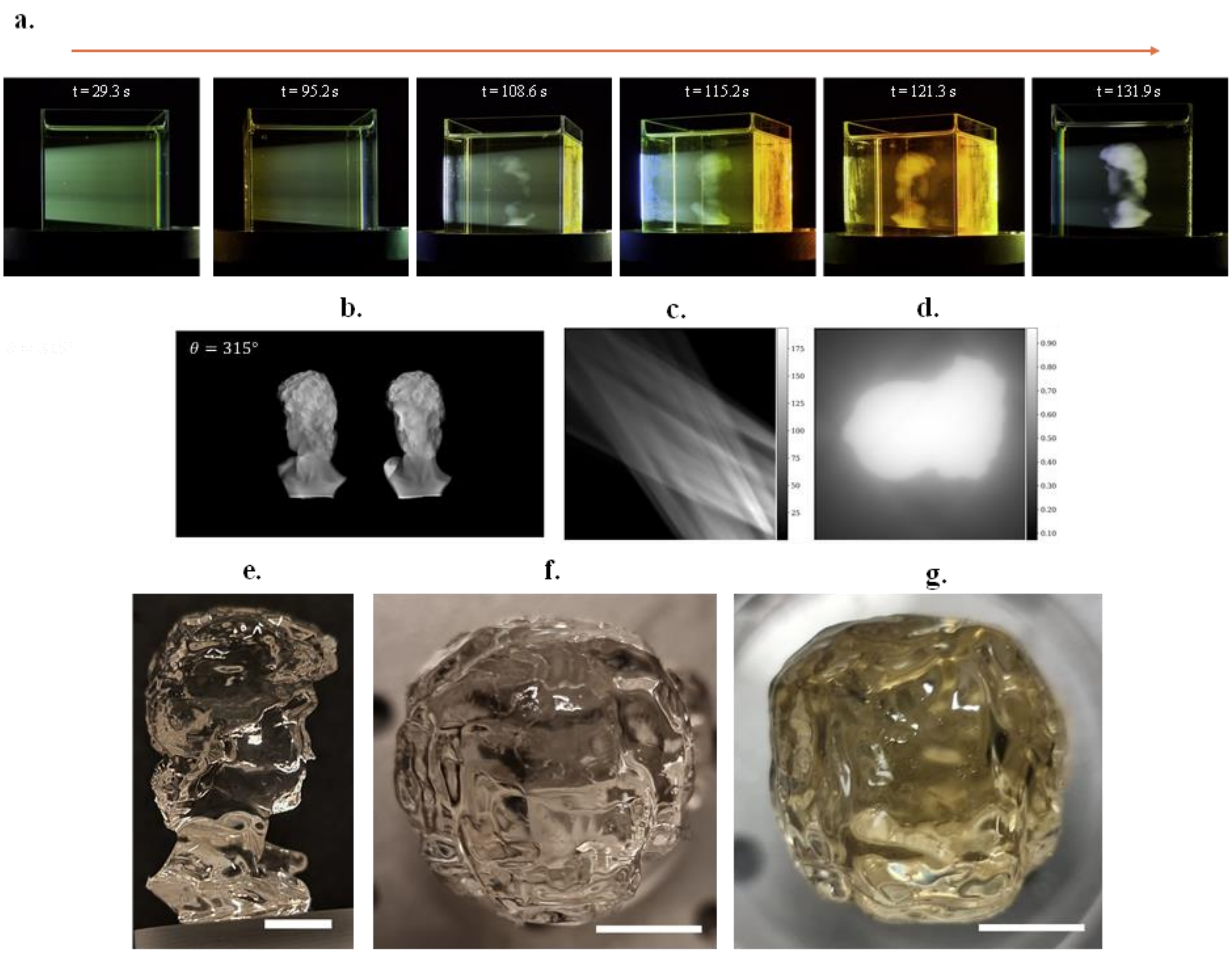

**Fig. 5: Printing results in a cubic container. (a)** Time-lapse sequence illustrating the photopolymerization process. **(b)** Sample projection image at 315°. **(c, d)** Computed 2D lateral cross-sectional dose distributions illustrating the resin exposure for **(c)** an individual projection and **(d)** the cumulative effect of all projections. The light propagation model accounts for both refraction at adjacent container faces and non-telecentric illumination. **(e)** Side view of the fully printed structure. **(f, g)** Comparative top views highlighting the influence of projection angle coverage: **(f)** represents the structure fabricated using the full range of projection angles, whereas **(g)** demonstrates the outcome when projections within ±5° near the edges are excluded, leading to increasingly rectilinear features. The scale bar is 10 mm.

## Large workpieces

To demonstrate the capabilities of our upscaled volumetric printing system, we fabricated three representative workpieces, one extended in the lateral dimensions, one vertically, and one both, shown in Fig. 6. Our first example is a life-sized human ear model measuring 56.5 × 23.7 × 70.9 mm³ (Fig. 6a). The model was oriented horizontally along its largest dimension and printed within a cylindrical container (OD = 125 mm, Movie S3). Due to the modest vertical dimension of the object, a single off-center projector sufficiently covered the entire lateral projection field (approximately 134 mm) during rotation. Therefore, only one projector was used for this design. Although the commercial projectors employed here did not allow direct hardware-based control over output intensity, we were able to modulate projection power by scaling pixel intensities in the computed patterns. To evaluate the impact of reduced illumination intensity on print quality and polymerization dynamics, the optimized pixel intensity values were deliberately scaled down by setting the absorption coefficient to 0.05. Under these conditions, the ear model required approximately 4 minutes to fully polymerize. In subsequent trials, we leveraged both projectors simultaneously, stitching their projections to enable the printing of vertically extended structures. Precise time synchronization was crucial to ensure the superposed beams delivered the correct dose into the resin. Any latency or desynchronization can shift the projections, compromising print fidelity. To achieve this, we used the GPU's clock to synchronize projection times, resulting in well-controlled alignment and accurate frame exposures. The second example (Fig. 6b) features a full-body sculpture of Francis of Assisi (National Gallery of Denmark), measuring approximately 25.1 × 72.2 × 19.8 $mm^3$. A cylindrical container with an outer diameter of around 50 mm was employed. For optimal attenuation/absorption, we used 0.05 wt.% CQ and 0.1 wt.% EDAB. The workpiece formed in under two minutes, capturing the sculpture's intricate surface details. Prolonged immersion in Isopropanol alcohol (IPA) during postprocessing produced a white coating on the workpiece, which, while unintentional, enhanced the visibility of its surface details but also caused a visible crack in the sculpture's head. For comparison, Fig. 6b also shows smaller workpieces produced in various scales using the same setup for comparison. During the experimental trials, the resin's low degree of conversion introduced several issues. Specifically, the neck region polymerized more slowly, sometimes compromising structural strength and causing the head to tilt post-print. Meanwhile, the sculpture's circular base often detached during extraction and post-processing. Mechanical drifts in the rotation stage also created dose inconsistencies at the boundary of the two projection zones, producing artifacts that affected local dimensional accuracy at the mid-section of the workpiece. However, we managed to minimize these defects via optimization of geometric arrangements in projection computation, without resorting to upgrading to customary optics. Lastly, we produced an upscaled head replica of Michelangelo's David, measuring approximately 42.6 × 78.2 × 50.1 $mm^3$ (Fig. 4c). Despite low photoinitiator concentration, the workpiece was produced in under three minutes. Similar artifacts at the intersection of the projector beams persisted and intensified modestly in larger containers, consistent with greater mechanical drift/wobble from looser tolerances (cf. the lateral streak on David's face, Fig. 6c).

## Discussion

We presented a methodology for scaling tomographic volumetric 3D printing by coordinating projections from multiple light sources and validated it using two off-the-shelf projectors with projection patterns computed via GPU-accelerated ray tracing. The results demonstrate large workpieces can be fabricated without loss of build rate in both cylindrical and non-cylindrical containers. By distributing the total irradiance across partially overlapping projectors, we effectively overcome the build volume limitations associated with single-projector setups. A key enabler is physics-aware pattern generation via ray tracing, which compensates for non-telecentricity, refraction, reflection, beam attenuation, and container geometry, thereby obviating custom telecentric optics or an index-matching vat.

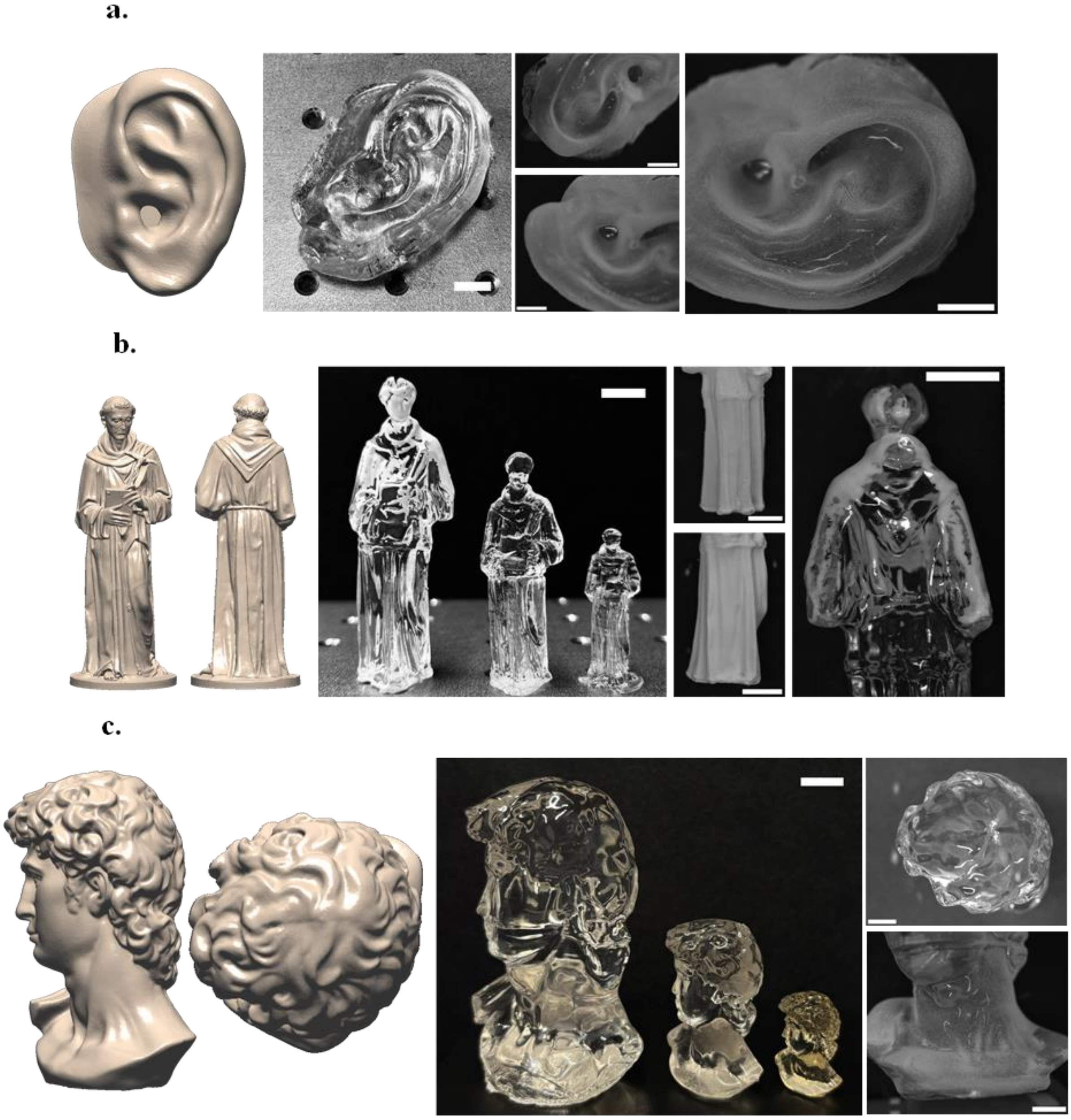

**Fig. 6: Scalable workpieces fabricated using dual-projector TVP setup: (a)** a life-size human ear model, **(b)** Frans af Assisi (National Gallery of Denmark, Copenhagen, Denmark), and **(c)** a replica of Michelangelo's David (National Gallery of Denmark, Copenhagen, Denmark). Scale bars are 10 mm.

Several challenges remain. Lowering the photoinitiator concentration was a pragmatic way to extend optical penetration beyond 60 mm, but it lowered the degree of conversion, resulting in production of soft parts that tore during extraction or in post-wash. Non-telecentric illumination and off-axis projector positioning also introduced beam overlaps, occasionally yielding overexposed regions despite dose optimization (see Fig. S2). At larger scales, container deformation, wall-thickness variation, and cumulative mechanical tolerances became more pronounced, underscoring the need for tighter container metrology and in-situ calibration. A recurring phenomenon near the end of polymerization was a transient upward motion of the printed part followed by sedimentation (most evident in larger samples) attributable to buoyancy driven by localized heating and volumetric expansion from exothermic reactions, consistent with photokinetic simulations (*23*, *24*). These findings emphasize the critical importance of precise timing and energy distribution optimization to achieve consistent curing across the entire print volume. Finally, feature fidelity was strongly limited by depth-of-focus: details near the focal plane (e.g., the surface of "Frans af Assisi") were well reproduced, whereas features farther from focus (e.g., the nose of "David") were incomplete, likely due to beam divergence. This trend aligns with our MTF measurements, which showed an increase in effective voxel size to ~700 µm at 40 mm away from the focal plane (Fig. S3).

## Materials and methods

### Experimental setup

Fig. S1a illustrates the dual-projector TVP setup used in this study. Two 4K UHD DLP laser projectors (BenQ LK954ST, 5100 ANSI lumens), labeled $P_1$ and $P_2$, were chosen for cost-effectiveness. Each projector's digital micromirror device (DMD) array natively supports 1920×1080 pixels, and rapid micromirror modulation enables an effective display of 3840×2160 pixels. Fig. S1b illustrates the layout of the DMD arrays within each projector. Every micromirror in the active region corresponds to a single pixel in the projected image. Surrounding this active area is the Pond of Micromirrors (POM), which contains partially functional micromirrors (*41*).

At the focal plane, each projector achieves a maximum projection area of approximately 92×52 mm$^2$, corresponding to theoretical (nominal) pixel pitches of around 24µm for the 4K resolution. The projectors are placed around a cylindrical container with an outer diameter (OD) of 125 mm (inner diameter (ID) 121 mm), with $P_2$ elevated to illuminate the upper sections of the container. A rotation stage, securely mounted at the center between the two projectors, holds the container in place via a custom 3D-printed holder (Fig. S1c). To enhance resolution at the focal plane and minimize optical aberrations, each projector is equipped with an achromatic doublet non-telecentric lens (Dia 50mm × 200 mm FL, TECHSPEC #45-179, Edmund Optics). Two CMOS cameras, labeled $C_1$ and $C_2$ (FLIR Grasshopper, Model GS3-U3-51S5M-C), provide lateral and vertical imaging for real-time monitoring of the photopolymerization process. These cameras also use adjustable working-distance, non-telecentric lenses (200 mm FL non-telecentric, TECHSPEC #45-179, Edmund Optics) to effectively relay the emitted light from the workpiece onto their sensors. Table S2 outlines the major system components.

**Optical alignment**

The projectors were initially aligned without the photoresin container, effectively stitching the lower and upper projection regions to form a combined projection area of approximately 138×104 $mm^2$ at the focal plane. Although the lateral dimension could have been extended to 184 mm, the presence of the POM region restricted further expansion. Once the photoresin container was introduced, additional lateral and vertical lensing distortions arose due to the non-telecentric optics, resulting in a visible gap in the 3D dose buildup. Ray tracing simulations were then performed to determine further offsets needed with container in position. Given limited available information on the internal projector design, each projector was modeled as a pinhole emitter with its unique resolution, color format, and geometry. Three parameters were critical for accurate raytracing simulations: (1) the 3D spatial coordinates of each virtual light source relative to the rotation center, (2) the native resolution of each projector, and (3) the angular field-of-view. The horizontal ($\approx \pm 23\ mm$) and vertical ($\approx \mp 26\ mm$) offsets provided the x and y coordinates for each virtual light source. To calculate the field-of-view angle and virtual source distance, projection dimensions were measured at two distinct intervals. Using trigonometric calculations, a vertical field-of-view angle ($\theta v$) of approximately 16.98° and a virtual source distance of roughly 177.1 mm were determined. Both projectors were modelled with their native 1920×1080-pixel DMD resolution despite their pseudo-4K capacity. These calculated parameters were integrated into the raytracer to simulate the collective dose distribution. Fig. S2 illustrates the effects of non-telecentricity and refraction on the 3D dose distribution for two distinct designs of varying dimensions, highlighting the importance of realignment. Containers matching each design's specific diameter were employed in the simulations. Simulation results indicated that lowering the upper projector (P2 in Fig. S1) by approximately 1.49 mm for the top model (container OD = 50mm) and 5.1 mm for the bottom model (container OD = 125mm), or shifting both projectors by half these distances, would effectively eliminate the gap in the dose distribution. Following these findings, the upper projector was adjusted downward accordingly, resulting in a reduced overall projection height of approximately 102.51 mm and 98.9 mm for the two cases, respectively.

**Optical resolution**

To quantify the optical resolution of the projectors, we used a CMOS camera (FLIR Grasshopper GS3-U3-51S5M-C) with a 3.45 µm pixel pitch (equivalent to 141 cycles/mm) and a native 2448×2048 pixel resolution. Slanted-edge test patterns were designed to match each projector's resolution specifications. Based on the projection geometry of the 4K projectors, the magnified pixel size (theoretical) was estimated to be 24 µm (corresponding to 20 cycles/mm). Each test pattern was projected individually, with the camera placed at the projector's focal plane to ensure maximum clarity. The captured images were processed in a separate Python script, which computed the Edge Spread Function (ESF) from the extracted edge data. After smoothing the ESF with a mean filter, the Line Spread Function (LSF) was obtained by numerical differentiation. Finally, the Modulation Transfer Function (MTF) was derived via Fourier transformation of the LSF, providing a measure of the projectors' spatial frequency response and overall optical resolution. Given the similarity of the projectors, MTF measurements were performed on one projector at various defocus positions. Fig. S3 displays the resulting MTF curves at the focal plane, demonstrating that spatial frequencies up to 27 cycles/mm were resolved with MTF values of at least 0.5, indicating the ability to reconstruct features as small as 19 μm near the focal plane. However, resolution degrades substantially with increased defocus: at ±10mm from the focal plane, the optical resolution widens to approximately 130 μm, and at ±40mm, it further expands to nearly 700 μm.

**Materials**

The photoresin was prepared by combining Bisphenol A glycerolate (1 glycerol/phenol) diacrylate (BPAGDA, CAS# 4687-94-9, Sigma–Aldrich) and poly (ethylene glycol) diacrylate (PEGDA, CAS# 26570-48-9, Sigma–Aldrich) at a ratio of 2:1. This ratio was chosen to ensure that the viscosity is large enough to keep workpieces stationary during printing. Camphorquinone (CQ, CAS# 10373-78-1, ≥96.5% purity, Sigma–Aldrich) and ethyl 4-dimethylaminobenzoate (EDAB, CAS# 10287-53-3, Sigma–Aldrich) were added as the photoinitiator and co-initiator, respectively.

**TomoPrint**

A software application, which we refer to as *TomoPrint*, was developed in Python, primarily using the PyOpenGL library, to generate the light patterns needed for multi-projector TVP. We defined a global coordinate system with its origin at the center of rotation and its y-axis aligned to the rotation axis, following the OpenGL convention (Fig. S1). All components, including the projectors, were positioned in this coordinate space. Custom GLSL (OpenGL Shading Language) shaders were implemented to accelerate raytracing on the GPU and accurately model the propagation of light through the components and materials used in the experimental setup. The ray-tracing framework comprised three main components: ray generation, ray traversal, and ray marching (both forward and inverse), each running on specialized compute shaders that leverage parallel operations. During ray traversal, intersection tests relied on either a mathematical representation for simple shapes (e.g., cylinders, cubes) or a polygon mesh (Fig. S4a). In the case of polygon meshes, a BVH structure accelerated ray-triangle intersections (Fig. S4b). For ray marching, the 3D design was voxelized either as a regular grid or an octree, depending on the available GPU memory (Fig. S4c). We adopted real-time conservative surface and solid voxelization methods of Schwarz and Seidel (*42*). A 3D texture typically stored the regular grid, an SSBO (Shader Storage Buffer Object) held the octree data, and a 2D texture array handled intermediate images (Fig. S3d).

To support real-time on-screen rendering, we integrated a rasterization pipeline alongside the compute pipeline, enabling efficient, real-time visualization of the evolving dose distribution. Vertical synchronization (VSync) aligned the rendered frames with the projector's refresh rate, preventing screen tearing and other visual artifacts. For repeatable experiments that required saving the rendered images, double-buffered Pixel Buffer Objects (PBOs) were employed to stream pixel data from the CPU to 2D textures on GPU. Each projector was assigned dedicated resources, such as full-screen quads, PBOs, textures, and rasterization shaders, to accommodate its specific resolution and color format. However, a shared OpenGL context ensured all screens could be managed and synchronized within a single rendering loop. The PBOs and textures were continuously swapped in sync with the projector's frame rate, ensuring the correct image index was displayed on each screen. Within this rendering loop, synchronization fences confirmed that all resources were fully loaded before any image was projected, signaling the GPU only after successful data transfer. This approach prevented partial or corrupted image displays, ultimately delivering high frame rates and accurate synchronization across all projectors.

**Dose optimization**

To ensure uniform dose distribution throughout the target volume and mitigate undesired dose accumulation in regions illuminated by multiple projectors, we used the object-space model optimization (OSMO) algorithm proposed by Rackson et al (*11*) to iteratively adjust the light patterns in each projector. In our calculations, we set the upper and lower dose thresholds to 0.9 and 0.7, respectively.

**Postprocessing**

After completion of the printing procedure, the test tube containing the fabricated workpiece along with the unpolymerized resin was heated to decrease the viscosity of the photoresin. This step facilitated the efficient removal of any residual resin. The excess material was carefully drained from the tube. Isopropanol alcohol (IPA) was then introduced into the test tube, which was subsequently sealed and gently shaken to ensure thorough cleaning of the workpiece's surface. The sealed tube was briefly exposed to ultraviolet light at 405 nm for 10 seconds to initiate further polymerization where necessary. Upon removing the workpiece from the tube, a hot air blower was used to evaporate any remaining IPA from its surface. The cleaned object was then placed in a controlled atmosphere saturated with nitrogen or argon gas and subjected to an extended UV curing process for 10 minutes to finalize the post-curing treatment.

**Spectrometry**

The emission spectra of the projector were captured using a Thorlabs CCS100 Compact Spectrometer (300–700 nm) equipped with a cosine corrector while the spectrometer was positioned at the projector's focal plane. To investigate the absorption properties of the photoresin, UV-Visible spectroscopy was performed using an Agilent G1103A spectrophotometer equipped with a 10 mm path length quartz cuvette. Absorbance spectra were collected across wavelengths ranging from 300 to 700 nm. All experimental procedures were carried out under dark conditions to prevent unintended photoreactions.

**Fourier Transform Infrared (FTIR) Spectroscopy**

Attenuated Total Reflectance–Fourier Transform Infrared (ATR-FTIR) spectroscopy was used to quantify the real-time double-bond conversion of resins using a PerkinElmer Spectrum 100 spectrometer (PerkinElmer, UK) equipped with a diamond ATR crystal. A circular resin holder with an internal diameter of 2 mm was positioned directly onto the ATR crystal surface to ensure consistent and reproducible sample placement. Four different initiator concentrations were tested (see Table S2), and 50 μL of each resin was placed into the holder to maintain a consistent volume. In-situ polymerization was initiated using a 470 nm blue LED (SOLIS-470C, Thorlabs) set at an intensity of 10 $mW/cm^2$. Light engine intensity was measured using a calibrated silicon photodiode power sensor (model S120VC, Thorlabs) connected to an optical power meter console (PM400, Thorlabs). FTIR spectra were recorded every 10 seconds within the range of 4000–600 $cm^{-1}$. Double-bond conversion was determined by monitoring the acrylate C=C stretching vibration at 1638 $cm^{-1}$, normalized to the aromatic ring at 1610 $cm^{-1}$ (see Fig S7).

**Viscosity**

The resin viscosity was measured using a Discovery HR-2 Hybrid rheometer (TA Instruments, USA) equipped with 40 mm diameter parallel plates. A constant plate gap of 1000 μm was set throughout the test. Measurements were carried out across a shear rate range of $10^{-3}$ to $10^{3}$ $s^{-1}$ at 25 °C on a temperature-controlled plate (see Fig. S8).

**Source of digital models**

The three-dimensional (3D) digital models utilized in this study were sourced from public online repositories. The STL file for the life-size human ear was acquired from the Thingiverse platform (https://www.thingiverse.com/). The digital models of the sculpture Saint Francis of Assisi and the replica of Michelangelo's David were obtained from the digital collection of the National Gallery of Denmark (SMK) (https://www.smk.dk/en/).

## References

1. B. E. Kelly, I. Bhattacharya, H. Heidari, M. Shusteff, C. M. Spadaccini, H. K. Taylor, Volumetric additive manufacturing via tomographic reconstruction. *Science (1979)* **363**, 1075–1079 (2019).

2. D. Loterie, P. Delrot, C. Moser, High-resolution tomographic volumetric additive manufacturing. *Nat Commun* **11** (2020).

3. M. Shusteff, A. E. M. Browar, B. E. Kelly, J. Henriksson, T. H. Weisgraber, R. M. Panas, N. X. Fang, C. M. Spadaccini, "One-step volumetric additive manufacturing of complex polymer structures" (2017); https://www.science.org.

4. B. E. Kelly, I. Bhattacharya, H. Heidari, M. Shusteff, C. M. Spadaccini, H. K. Taylor, Volumetric additive manufacturing via tomographic reconstruction. *Science (1979)* **363**, 1075–1079 (2019).

5. P. N. Bernal, M. Bouwmeester, J. Madrid-Wolff, M. Falandt, S. Florczak, N. G. Rodriguez, Y. Li, G. Größbacher, R. A. Samsom, M. van Wolferen, L. J. W. van der Laan, P. Delrot, D. Loterie, J. Malda, C. Moser, B. Spee, R. Levato, Volumetric Bioprinting of Organoids and Optically Tuned Hydrogels to Build Liver-Like Metabolic Biofactories. *Advanced Materials* **34** (2022).

6. A. Boniface, F. Maître, J. Madrid-Wolff, C. Moser, Volumetric helical additive manufacturing. *Light: Advanced Manufacturing* **4**, 124–132 (2023).

7. J. Madrid-Wolff, J. Toombs, R. Rizzo, P. N. Bernal, D. Porcincula, R. Walton, B. Wang, F. Kotz-Helmer, Y. Yang, D. Kaplan, Y. S. Zhang, M. Zenobi-Wong, R. R. McLeod, B. Rapp, J. Schwartz, M. Shusteff, H. Talyor, R. Levato, C. Moser, A review of materials used in tomographic volumetric additive manufacturing. Springer Nature [Preprint] (2023). https://doi.org/10.1557/s43579-023-00447-x.

8. I. Bhattacharya, J. Toombs, H. Taylor, High fidelity volumetric additive manufacturing. *Addit Manuf* **47** (2021).

9. J. Toombs, C. C. Li, H. Taylor, "Roll-to-roll tomographic volumetric additive manufacturing for continuous production of microstructures on long flexible substrates" (2024); https://arxiv.org/abs/2402.10955.

10. J. T. Toombs, I. K. Shan, H. K. Taylor, Ethyl Cellulose-Based Thermoreversible Organogel Photoresist for Sedimentation-Free Volumetric Additive Manufacturing. *Macromol Rapid Commun* **44** (2023).

11. C. M. Rackson, K. M. Champley, J. T. Toombs, E. J. Fong, V. Bansal, H. K. Taylor, M. Shusteff, R. R. McLeod, Object-space optimization of tomographic reconstructions for additive manufacturing. *Addit Manuf* **48** (2021).

12. T. Chen, H. Li, X. Liu, Statistical iterative pattern generation in volumetric additive manufacturing based on ML-EM. *Opt Commun* **537**, 129448 (2023).

13. C. C. Li, J. Toombs, H. K. Taylor, T. J. Wallin, Tomographic projection optimization for volumetric additive manufacturing with general band constraint Lp-norm minimization. *Addit Manuf* **94**, 104447 (2024).

14. A. Orth, K. L. Sampson, K. Ting, J. Boisvert, C. Paquet, Correcting ray distortion in tomographic additive manufacturing. *Opt Express* **29**, 11037 (2021).

15. D. Webber, Y. Zhang, M. Picard, J. Boisvert, C. Paquet, A. Orth, Versatile volumetric additive manufacturing with 3D ray tracing. *Opt Express* **31**, 5531 (2023).

16. B. Nicolet, F. Wechsler, J. Madrid-Wolff, C. Moser, W. Jakob, Inverse Rendering for Tomographic Volumetric Additive Manufacturing. *ACM Trans Graph* **43** (2024).

17. W. Jakob, S. Speierer, N. Roussel, M. Nimier-David, D. Vicini, B. Nicolet, M. Crespo, V. Leroy, Z. Zhang, Mitsuba 3 renderer. https://mitsuba-renderer.org 3.1.1 [Preprint] (2022).

18. J. Madrid-Wolff, A. Boniface, D. Loterie, P. Delrot, C. Moser, Controlling Light in Scattering Materials for Volumetric Additive Manufacturing. *Advanced Science* **9**, 2105144 (2022).

19. A. Orth, K. L. Sampson, Y. Zhang, K. Ting, D. A. van Egmond, K. Laqua, T. Lacelle, D. Webber, D. Fatehi, J. Boisvert, C. Paquet, On-the-fly 3D metrology of volumetric additive manufacturing. *Addit Manuf* **56** (2022).

20. P. N. Bernal, P. Delrot, D. Loterie, Y. Li, J. Malda, C. Moser, R. Levato, Volumetric Bioprinting of Complex Living-Tissue Constructs within Seconds. *Advanced Materials* **31** (2019).

21. J. T. Toombs, M. Luitz, C. C. Cook, S. Jenne, C. C. Li, B. E. Rapp, F. Kotz-Helmer, H. K. Taylor, Volumetric additive manufacturing of silica glass with microscale computed axial lithography. *Science (1979)* **376**, 308–312 (2022).

22. B. Wang, E. Engay, P. R. Stubbe, S. Z. Moghaddam, E. Thormann, K. Almdal, A. Islam, Y. Yang, Stiffness control in dual color tomographic volumetric 3D printing. *Nat Commun* **13** (2022).

23. R. Salajeghe, B. Šeta, N. Pellizzon, C. G. S. Kruse, D. Marla, A. Islam, J. Spangenberg, Numerical modeling of tomographic volumetric additive manufacturing based on energy threshold method. *Addit Manuf* **96**, 104552 (2024).

24. T. H. Weisgraber, M. P. de Beer, S. Huang, J. J. Karnes, C. C. Cook, M. Shusteff, Virtual Volumetric Additive Manufacturing (VirtualVAM). *Adv Mater Technol* **8** (2023).

25. A. Orth, D. Webber, Y. Zhang, K. L. Sampson, H. W. de Haan, T. Lacelle, R. Lam, D. Solis, S. Dayanandan, T. Waddell, T. Lewis, H. K. Taylor, J. Boisvert, C. Paquet, Deconvolution volumetric additive manufacturing. *Nat Commun* **14** (2023).

26. C. C. Cook, E. J. Fong, J. J. Schwartz, D. H. Porcincula, A. C. Kaczmarek, J. S. Oakdale, B. D. Moran, K. M. Champley, C. M. Rackson, A. Muralidharan, R. R. McLeod, M.

Shusteff, Highly Tunable Thiol-Ene Photoresins for Volumetric Additive Manufacturing. *Advanced Materials* **32** (2020).

27. T. Chen, S. You, L. Xu, C. Cao, H. Li, C. Kuang, X. Liu, High-fidelity tomographic additive manufacturing for large-volume and high-attenuation situations using expectation maximization algorithm. *Addit Manuf* **80** (2024).

28. M. Shusteff, A. E. M. Browar, B. E. Kelly, J. Henriksson, T. H. Weisgraber, R. M. Panas, N. X. Fang, C. M. Spadaccini, One-step volumetric additive manufacturing of complex polymer structures. *Sci Adv* **3**, eaao5496 (2017).

29. R. Rizzo, D. Ruetsche, H. Liu, M. Zenobi-Wong, Optimized Photoclick (Bio)Resins for Fast Volumetric Bioprinting. *Advanced Materials* **33** (2021).

30. M. Kollep, G. Konstantinou, J. Madrid-Wolff, A. Boniface, L. Hagelüken, P. V. W. Sasikumar, G. Blugan, P. Delrot, D. Loterie, J. Brugger, C. Moser, Tomographic Volumetric Additive Manufacturing of Silicon Oxycarbide Ceramics. *Adv Eng Mater* **24** (2022).

31. C. M. Rackson, J. T. Toombs, M. P. De Beer, C. C. Cook, M. Shusteff, H. K. Taylor, R. R. McLeod, Latent image volumetric additive manufacturing. *Opt Lett* **47**, 1279 (2022).

32. M. Falandt, P. N. Bernal, O. Dudaryeva, S. Florczak, G. Größbacher, M. Schweiger, A. Longoni, C. Greant, M. Assunção, O. Nijssen, S. van Vlierberghe, J. Malda, T. Vermonden, R. Levato, Spatial-Selective Volumetric 4D Printing and Single-Photon Grafting of Biomolecules within Centimeter-Scale Hydrogels via Tomographic Manufacturing. *Adv Mater Technol* **8** (2023).

33. Q. Thijssen, A. Quaak, J. Toombs, E. De Vlieghere, L. Parmentier, H. Taylor, S. Van Vlierberghe, Volumetric Printing of Thiol-Ene Photo-Cross-Linkable Poly(ε-caprolactone): A Tunable Material Platform Serving Biomedical Applications. *Advanced Materials*, doi: 10.1002/adma.202210136 (2023).

34. P. Chansoria, D. Rütsche, A. Wang, H. Liu, D. D'Angella, R. Rizzo, A. Hasenauer, P. Weber, W. Qiu, N. B. M. Ibrahim, N. Korshunova, X. H. Qin, M. Zenobi-Wong, Synergizing Algorithmic Design, Photoclick Chemistry and Multi-Material Volumetric Printing for Accelerating Complex Shape Engineering. *Advanced Science* **10** (2023).

35. M. Xie, L. Lian, X. Mu, Z. Luo, C. E. Garciamendez-Mijares, Z. Zhang, A. López, J. Manríquez, X. Kuang, J. Wu, J. K. Sahoo, F. Z. González, G. Li, G. Tang, S. Maharjan, J. Guo, D. L. Kaplan, Y. S. Zhang, Volumetric additive manufacturing of pristine silk-based (bio)inks. *Nat Commun* **14** (2023).

36. L. Barbera, J. Madrid-Wolff, R. Emma, K. Masania, A. Boniface, D. Loterie, P. Delrot, C. Moser, A. R. Studart, Multimaterial Volumetric Printing of Silica-Based Glasses. *Adv Mater Technol* **9** (2024).

37. L. Lian, M. Xie, Z. Luo, Z. Zhang, S. Maharjan, X. Mu, C. E. Garciamendez-Mijares, X. Kuang, J. K. Sahoo, G. Tang, G. Li, D. Wang, J. Guo, F. Z. González, V. Abril

Manjarrez Rivera, L. Cai, X. Mei, D. L. Kaplan, Y. S. Zhang, Rapid Volumetric Bioprinting of Decellularized Extracellular Matrix Bioinks (Adv. Mater. 34/2024). *Advanced Materials* **36**, 2470274 (2024).

38. M. Pharr, G. Humphreys, *Physically Based Rendering, Second Edition: From Theory To Implementation* (Morgan Kaufmann Publishers Inc., San Francisco, CA, USA, ed. 2nd, 2010).

39. S. Chandrasekhar, *Radiative Transfer* (Dover Publication, New York, 1960).

40. J. Amanatides, A. Woo, "A Fast Voxel Traversal Algorithm for Ray Tracing" in *Eurographics* (1987).

41. Texas Instruments, "DLP471TE .47 HSSI DMD Datasheet (Rev. B)" (2022); https://www.ti.com/product/DLP471TE.

42. M. Schwarz, H.-P. Seidel, "Fast Parallel Surface and Solid Voxelization on GPUs" (2010); https://doi.org/http://doi.acm.org/10.1145/1882261.1866201.

Supplementary information for

# Upscaling Tomographic Volumetric 3D Printing via Virtually Stitching of Coordinated Projections

Hossein Safari[1,2], S. Kaveh Hedayati[3], Aminul Islam[3], Yi Yang[1,2,4*]

[1] Department of Chemistry, Technical University of Denmark; 2800 Kongens Lyngby, Denmark.

[2] Center for Energy Resources Engineering, Technical University of Denmark; 2800 Kongens Lyngby, Denmark.

[3] Department of Civil and Mechanical Engineering, Technical University of Denmark; 2800 Kongens Lyngby, Denmark.

[4] PERFI Technologies ApS; Kemitorvet 207, 2800 Kongens Lyngby, Denmark

* To whom correspondence should be addressed (email: yi@perfi.dk)

**This Supplementary Information contains**

9 supplementary figures (Figs S1 – S9)

5 supplementary movies (Mov S1 – S5)

3 supplementary tables (Tables S1 – S3)

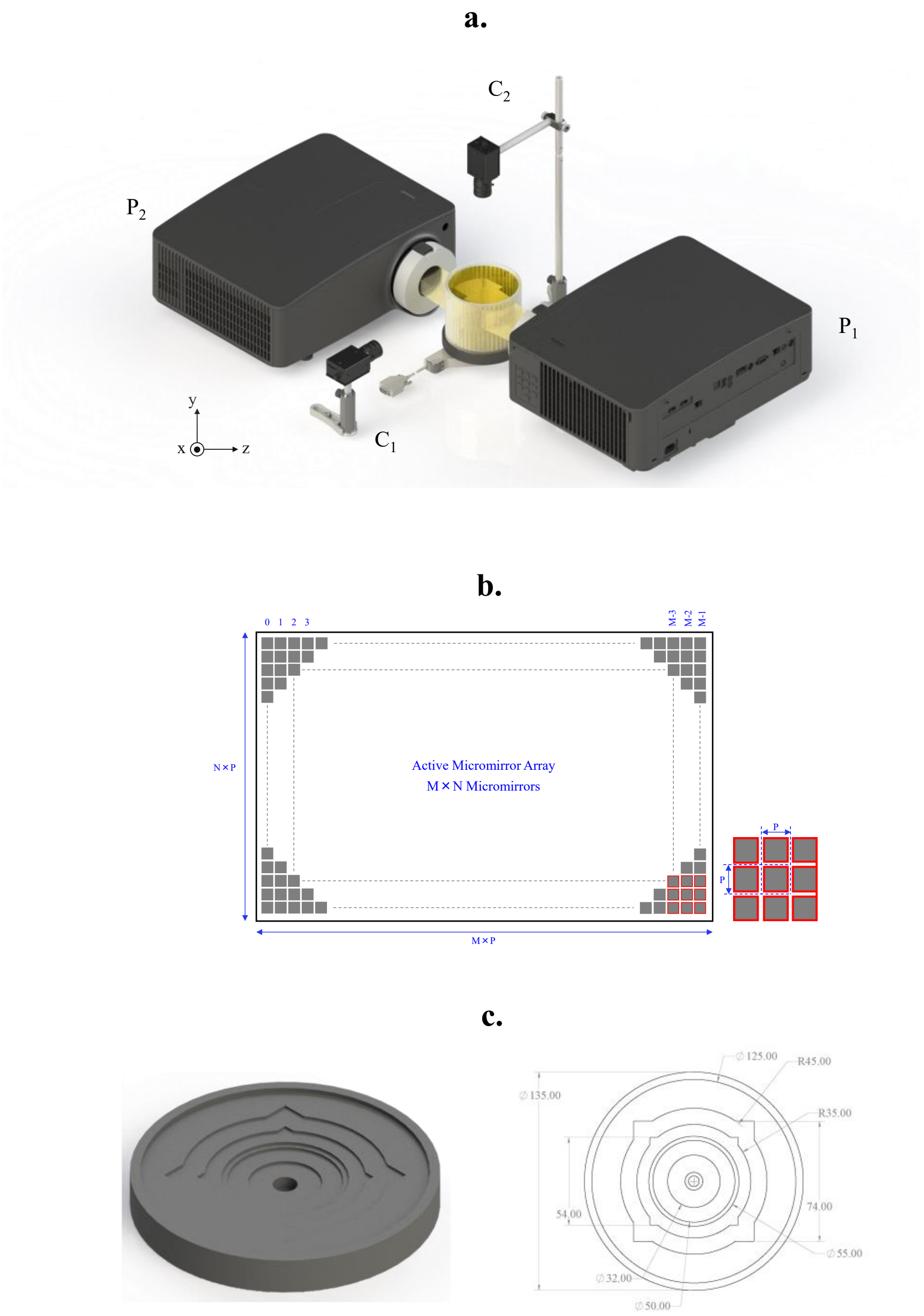

**Fig. S1: (a)** Schematic design of the dual-projector setup used in this study. Each projector uses a non-telecentric lens ($NTL_1$ and $NTL_2$), characterized by focal lengths $f_1$ and $f_2$, respectively. Major system components are outlined in Table S1. **(b)** Layout of the micromirror arrays in the projectors, where $M$ and $N$ denote the horizontal and vertical dimensions, and $P$ is the pixel pitch (5.4 µm). The gray-shaded Pond of Micromirrors (POM) region contains partially active micromirrors. **(c)** CAD model of the 3D-printed holder, along with its key design parameters.

**Table S1: Components used in the dual-projector TVP setup depicted in Fig. S1**

| Identifier | Component |
|---|---|
| **$P_1$, $P_2$** | BenQ LK954ST 5100 ANSI Lumen 4K |
| $C_1$, $C_2$ | FLIR Grasshopper GS3-U3-51S5M-C |
| $NTL_1$, $NTL_2$ | 200 mm FL, TECHSPEC #45-179, Edmund Optics |

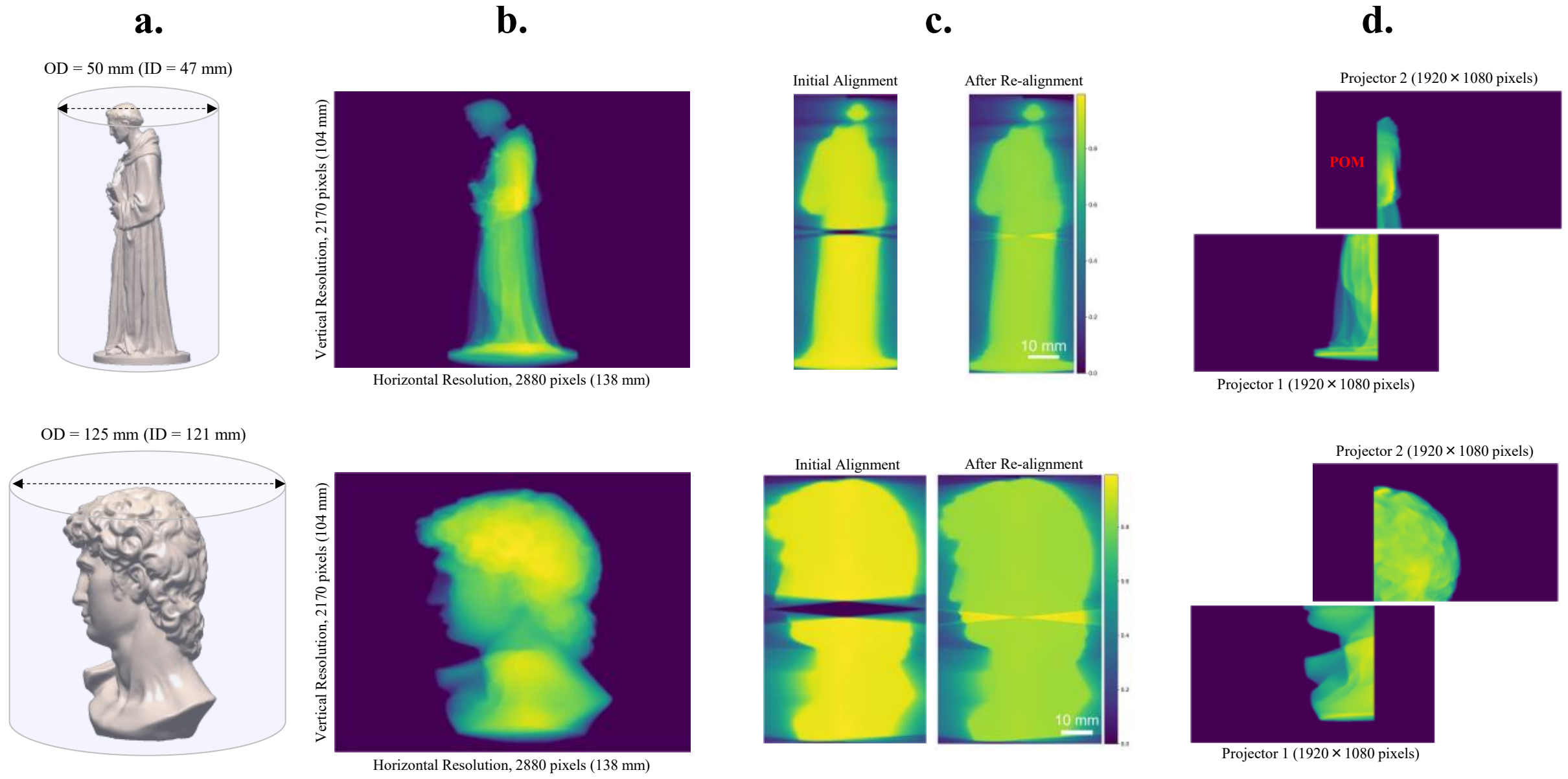

**Fig. S2: Effect of non-telecentric optics and refraction on projector alignment. (a)** Illustrations of two 3D designs placed in cylindrical containers sized to match their dimensions. **(b)** Representative projections computed for each design, assuming a hypothetical projector with a large field of view. **(c)** Central-slice comparison of optimized 3D dose distributions before and after projector realignment, highlighting a dose gap resulting from neglecting vertical lensing effects during initial alignment. **(d)** Projector-specific projections after optimization of pixel intensities. The scale bars represent 10 mm.

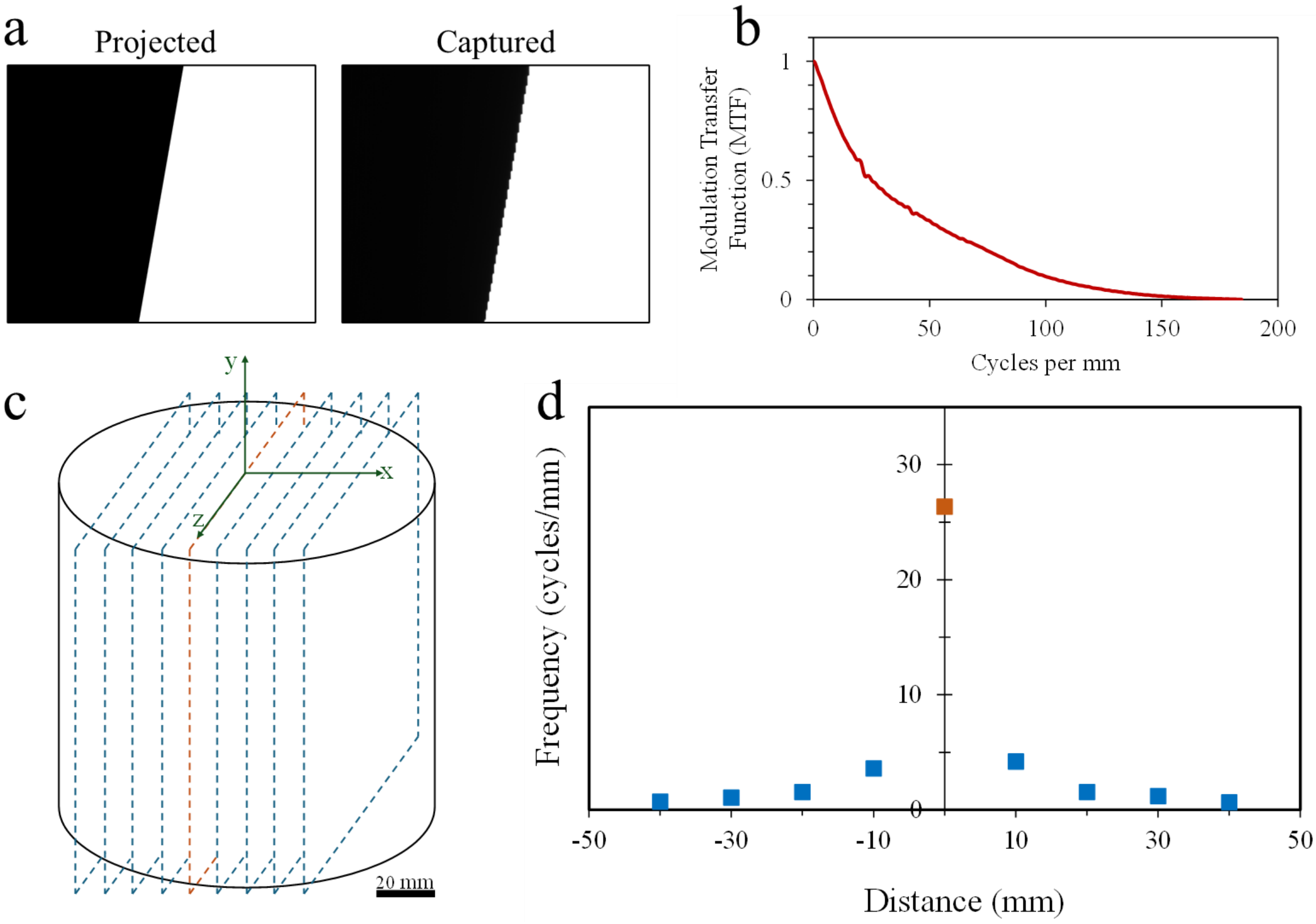

**Fig. S3: Optical characteristics of the projectors. (a)** Projected slant-edge image at 3840×2160 resolution and its corresponding captured image. **(b)** MTF results at the focal plane. **(c)** Schematic illustrating various measurement distances relative to the focal plane. **(d)** MTF measurements obtained at different axial planes.

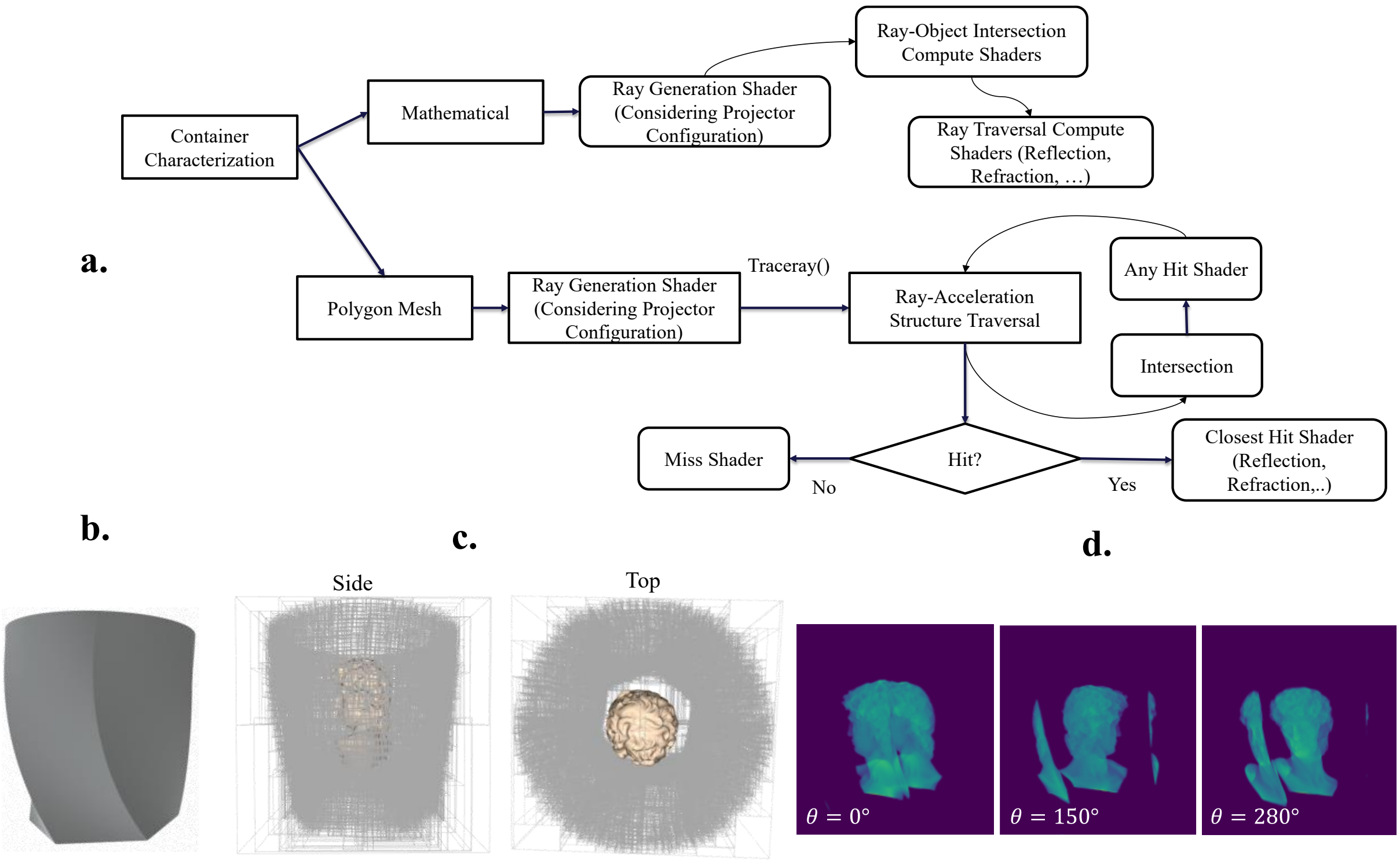

**Fig. S4: Ray tracing in arbitrary-shaped containers. (a)** A workflow driven by container characterization for ray generation and traversal. **(b)** An arbitrary container and **(c)** side and top views of its BVH structure embedding a 3D design intended for fabrication. **(d)** An example of the final rendered projection.

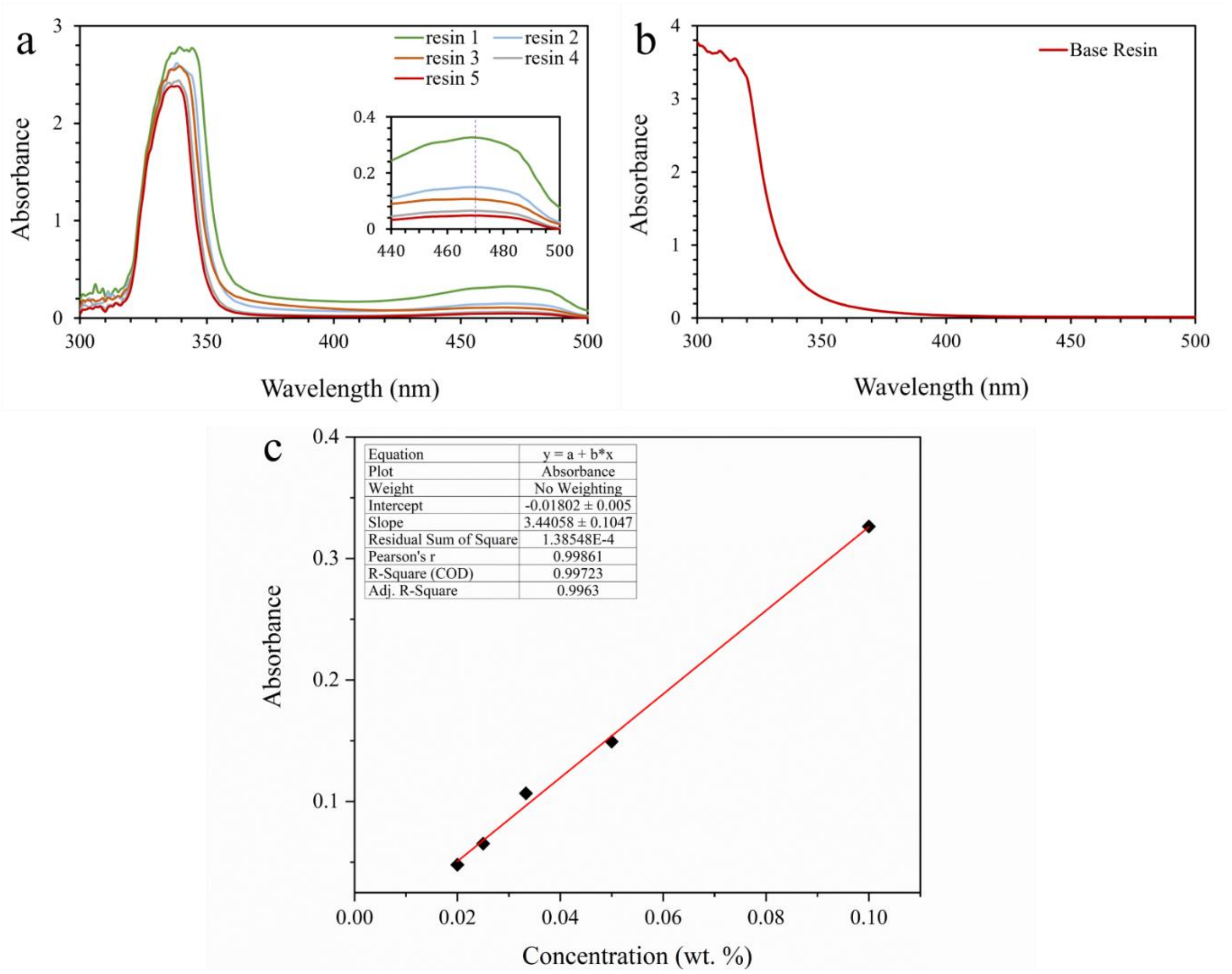

**Fig. S5: UV-Vis spectroscopy of the resins. (a)** Absorption profile at different PI concentrations. **(b)** Absorption profile of the Base resin without any PI. **(c)** Calculation of extinction coefficient of PIs at 470 nm.

**Table S2: Photoinitiator concentrations (CQ/EDAB) with corresponding peak absorbance, depth of penetration ($D_p$), and molar absorptivity in resin formulations**

| Resin | CQ [wt. %] | EDAB [wt. %] | Peak Absorbance | $D_p$ [cm] |
|---|---|---|---|---|
| **1** | 0.1 | 0.2 | 0.33 | 1.33 |
| **2** | 0.05 | 0.1 | 0.15 | 2.90 |
| **3** | 0.033 | 0.067 | 0.11 | 4.03 |
| **4** | 0.025 | 0.05 | 0.067 | 6.58 |
| **5** | 0.02 | 0.04 | 0.049 | 8.95 |

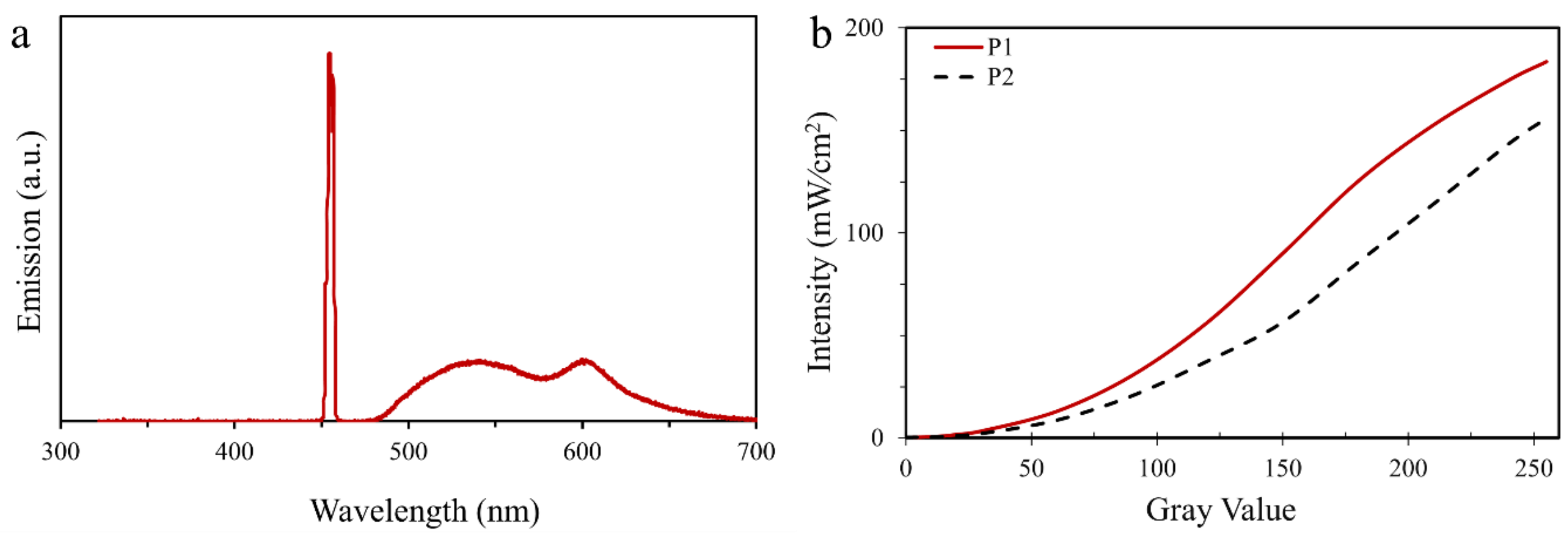

**Fig. S6: Projectors specifications: (a)** Emission profile, displayed for only one projector due to their near-identical characteristics. **(b)** Intensity profile of both projectors versus grayscale values.

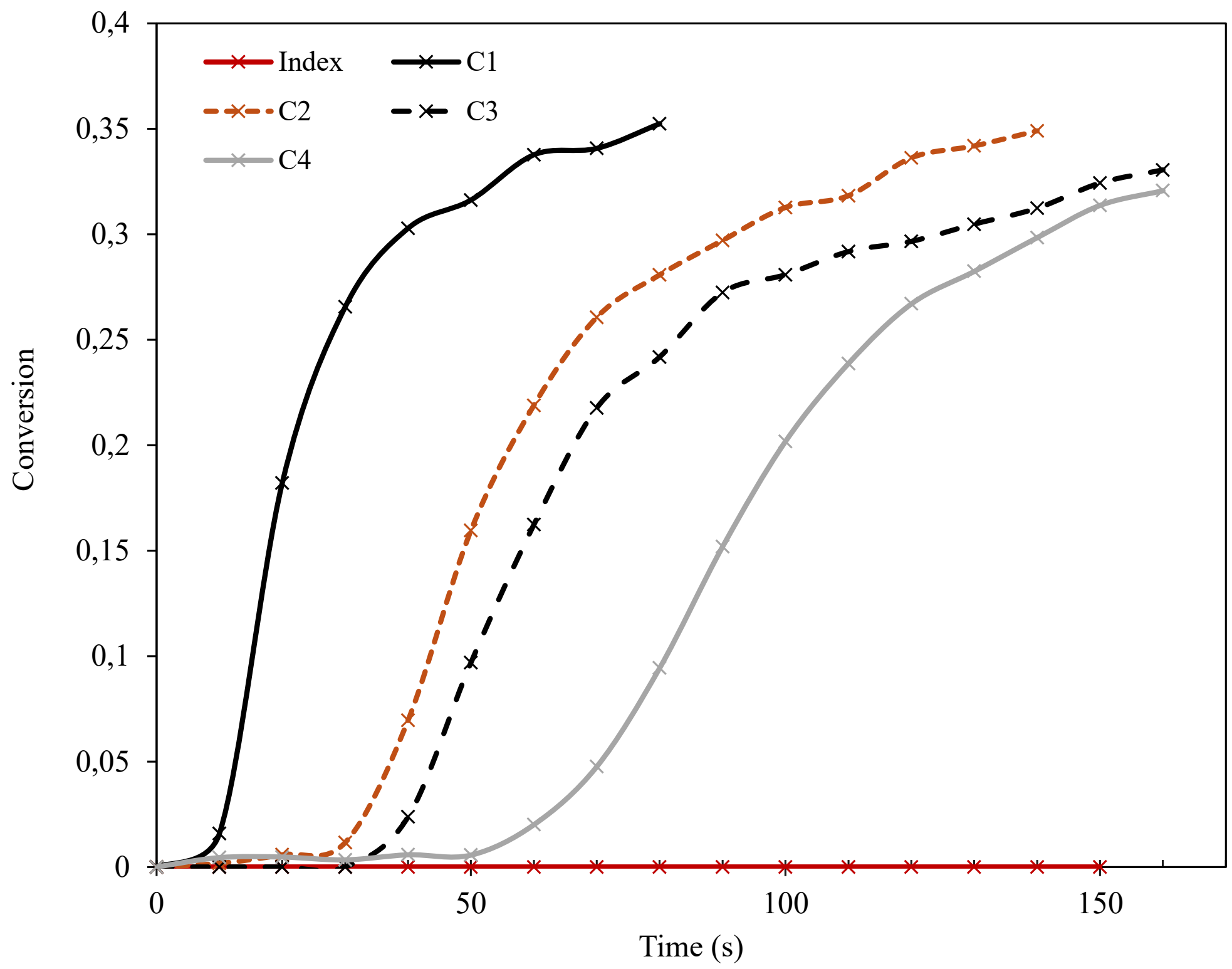

**Fig. S7: Double bond conversion of the first four resins provided in Table S2 as well as the index liquid (without photoinitiator).** The irradiance of the visible light was 10 mW/cm².

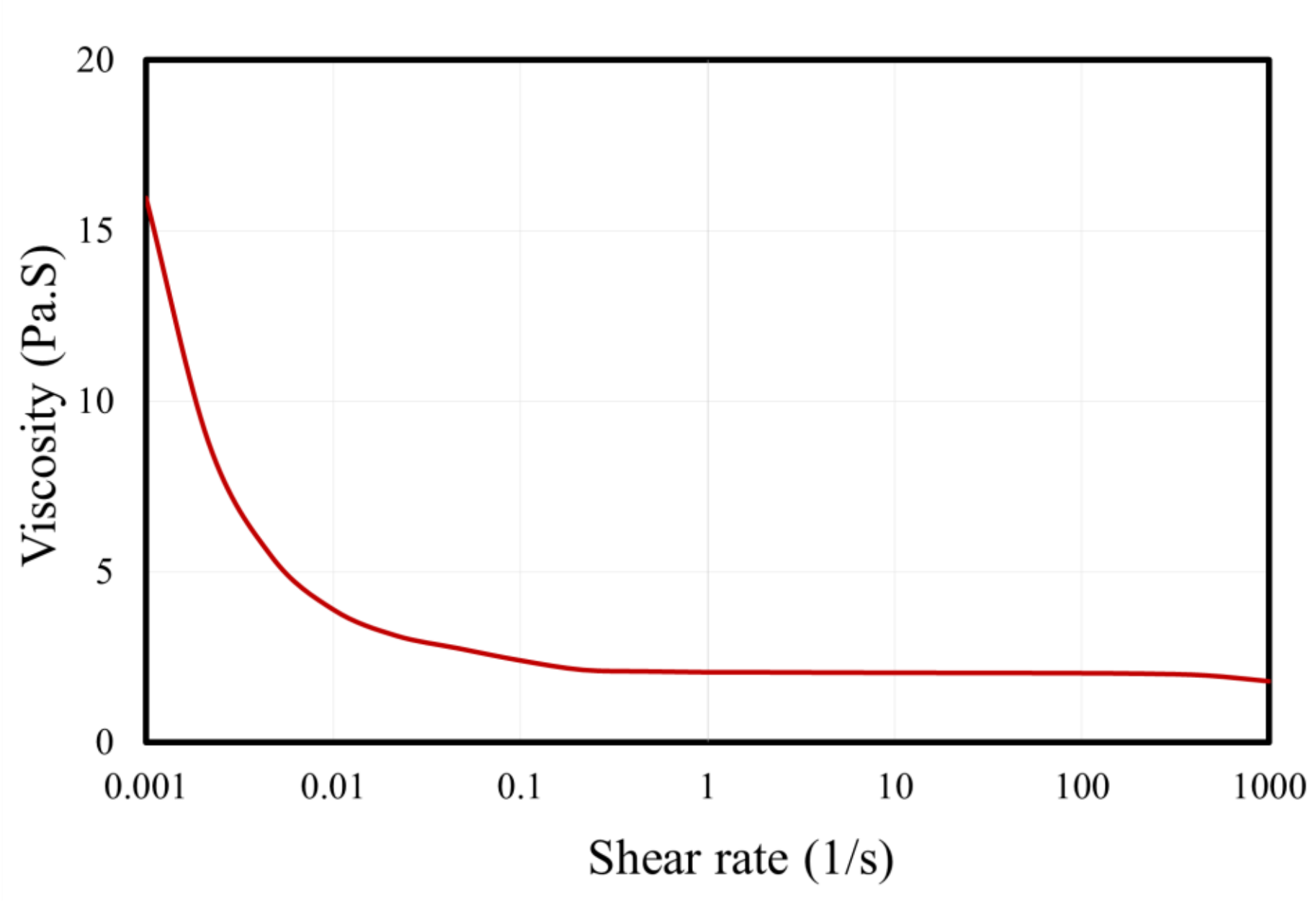

**Fig. S8: Dynamic viscosity of the photoreactive resin.** The dynamic viscosity of the used material was measured using a digital rheometer at room temperature, and the plot of viscosity versus shear rate is shown.

**Supplementary Movies**

**Mov. S1: Pyramid.** Top view of the printing process of a pyramid shaped phantom (≈ $40 \times 16 \times 40$ mm$^3$) inside a cylindrical container with an outer diameter of 125 mm using one projector.

**Mov. S2: David in cube.** Side view recording of the printing process of a replica of Michelangelo's David (National Gallery of Denmark) (≈32 × 52.5 × 31.8 mm$^3$) inside a cubic container with an edge length of 74 mm using one projector.

**Mov. S3: Ear.** Top view of the printing process of a life-sized human ear model (≈56.5 × 23.7 × 70.9 mm$^3$) inside a cylindrical container of outer diameter 125 mm using one projector.

**Mov. S4: Frans af assisi.** Side view of the printing process of a full-body sculpture of Francis of Assisi (National Gallery of Denmark) (≈25.1 × 72.2 × 19.8 mm$^3$) inside a cylindrical container of outer diameter 50 mm using two projectors.

**Mov. S5: David.** Side view recording of the printing process of an upscaled replica of Michelangelo's David (≈42.6 × 78.2 × 50.1 mm$^3$) inside a cylindrical container of outer diameter 125 mm using two projectors.